**Quantum phase transitions in two-dimensional superconductors: a review on recent experimental progress**

Ziqiao Wang[1,6], Yi Liu[2,3,6], Chengcheng Ji[1,5] and Jian Wang[1,4,5*]

[1] International Center for Quantum Materials, School of Physics, Peking University, Beijing 100871, China

[2] Department of Physics and Beijing Key Laboratory of Opto-electronic Functional Materials & Micro-nano Devices, Renmin University of China, Beijing 100872, China

[3] Key Laboratory of Quantum State Construction and Manipulation (Ministry of Education), Renmin University of China, Beijing 100872, China

[4] Collaborative Innovation Center of Quantum Matter, Beijing 100871, China

[5] Hefei National Laboratory, Hefei 230088, China

[6] These authors contributed equally to this work.

[*] Authors to whom any correspondence should be addressed.

E-mail: jianwangphysics@pku.edu.cn

**Abstract**

Superconductor-insulator/metal transition (SIT/SMT) as a paradigm of quantum phase transition has been a research highlight over the last three decades. Benefit from recent developments in the fabrication and measurements of two-dimensional (2D) superconducting films and nanodevices, unprecedented quantum phenomena have been revealed in the quantum phase transitions of 2D superconductors. In this review, we introduce the recent progress on quantum phase transitions in 2D superconductors, focusing on the quantum Griffiths singularity (QGS) and anomalous metal state. Characterized by a divergent critical exponent when approaching zero temperature, QGS of SMT is discovered in ultrathin crystalline Ga films and subsequently detected in various 2D superconductors. The universality of QGS indicates the profound influence of quenched disorder on quantum phase transitions. Besides, in a 2D superconducting system, whether a metallic ground state can exist is a long-sought mystery. Early experimental studies indicate an intermediate metallic state in the quantum phase transition of 2D superconductors. Recently, in high-temperature superconducting films with patterned nanopores, a robust anomalous metal state (*i.e.*, quantum metal or Bose metal) has been detected, featured as the saturated resistance in the low temperature regime. Moreover, the charge-2$e$ quantum oscillations are observed in nanopatterned films, indicating the bosonic nature of the anomalous metal state and ending the debate on whether bosons can exist as a metal. The evidences of the anomalous metal states have also been reported in crystalline epitaxial thin films and exfoliated nanoflakes, as well as granular composite films. High quality filters are used in these works to exclude the influence of external high frequency noises in ultralow temperature measurements. The observations of QGS and metallic ground states in 2D superconductors not only reveal the prominent role of quantum fluctuations and dissipations but also provide new perspective to explore quantum phase transitions in superconducting systems.

Keywords: quantum phase transition, quantum Griffiths singularity, anomalous metal state, two-dimensional superconductor

## 1. Introduction

As a paradigm of quantum phase transition, superconductor-insulator/metal transition (SIT/SMT) in 2D superconductors provides an ideal platform to study the quantum criticality, where the disorder and quantum fluctuations have dramatic effects on the quantum behavior. Over the last three decades, the study of SIT/SMT has been a research highlight in both experimental and theoretical condensed matter physics [1-15]. The 2015 Oliver E. Buckley Condensed Matter Physics Prize was awarded to Aharon Kapitulnik, Allen Goldman, Arthur Hebard, and Matthew Fisher for their discovery and pioneering investigations of the SIT. In Chapter 1 of this review, we start out by reviewing representative works on SIT/SMT, followed by basic concepts of quantum phase transition and quantum criticality.

The superconducting state is characterized by a complex order parameter $\Psi = \Delta e^{i\phi}$, where $\Delta$ and $\phi$ are amplitude and phase of the order parameter. The zero-resistance superconducting state is realized in a system with nonzero amplitude and long-range phase coherence. The suppression of superconductivity is thus attributed to either the suppression of the amplitude or the loss of long-range phase coherence [6, 15]. The former mechanism, known as the fermionic scenario, suggests that the breaking of Cooper pairs could occur when the disorder enhanced Coulomb interaction surpasses the phonon-mediated attractive interaction of Cooper pairs [5]. Fisher proposed an alternative mechanism that, in the presence of phase fluctuation, the phase coherence of the superconductivity could be destroyed even with a nonzero amplitude [3], known as the bosonic scenario. Another major outcome of this work [3] is the development of scaling theory that describes the scaling dependence of the resistance in the quantum critical region.

Applying high external magnetic field is a common method to induce superconducting phase transitions. The Cooper pairs can be destroyed by the external magnetic field via the orbital or spin pair-breaking effects. The orbital pair-breaking effect originates from the coupling between the magnetic field and the electron momentum. With increasing magnetic field, the Lorentz force exceeds the binding force of the electrons and thus breaks the Cooper pairs. In 2D superconductors, the orbital pair-breaking effect is eliminated under parallel magnetic field and the in-plane upper critical field is determined by the spin pair-breaking effect. For conventional *s*-wave superconductors, the Cooper pair breaks when the spin paramagnetic energy surpasses the binding energy of the Cooper pair, known as the Pauli limit or the Clogston-Chandrasekhar limit [16, 17]. It has been shown that the pair-breaking effects play important roles in the magnetic field driven SMT and the quantum critical behaviors can be understood in the framework of pair-breaking quantum phase transitions [18-20]. More interestingly, a divergent dynamical critical exponent has been revealed when approaching the quantum critical point in atomically thin crystalline Ga films [14], which marks the experimental discovery of quantum Griffiths singularity (QGS) in 2D superconductors and demonstrates the dramatic effect of quenched disorder on SMT. In Chapter 2, we will focus on the experimental progress on QGS in superconducting systems. The theoretical developments of QGS and the experimental signatures of QGS in ferromagnetic quantum phase transitions will also be briefly introduced.

Besides the pair-breaking effect, the suppression of superconductivity may also have its origin from the classical motion or quantum tunneling of vortices due to the loss of phase coherence. A representative example is the observation of Berezinskii-Kosterlitz-Thouless (BKT) transition in 2D superconducting systems [21-23]. Above the BKT transition temperature, the vortex-antivortex pairs break up into individual vortices, and the phase fluctuations destroy the global phase coherence leading to the breakdown of superconductivity [24]. Besides, the magnetic field penetrates into the type-II superconductors above the lower critical field, forming a regular array of vortices (i.e., the Abrikosov vortex lattice [25]). In the absence of pinning effect, the Lorentz force induced by the magnetic field drives the vortices to move and hence results in non-zero resistance when the current is applied [26, 27]. In real materials, however, the defects can pin the vortices so that zero resistance persists until the critical current or the critical field is reached. Superconducting vortices and pinning centers have been demonstrated to be associated with disorder driven phase transitions and dynamic instabilities in 2D superconducting systems [28-31]. Furthermore, the quantum tunneling of vortices could give rise to non-zero resistance in a superconducting system. In the absence of dissipation, the SIT occurs when the quantum tunneling of vortices becomes predominant and the strong phase fluctuations prevent the system forming long-range phase coherence [24]. On the other hand, the coupling to quasiparticle excitations leads to the ohmic dissipation, which significantly changes the vortices dynamics and further gives rise to the experimental features of the anomalous metal state [32]. The experimental progress and the physical understandings on the anomalous metal state will be illustrated in detail in Chapter 3.

In recent decades, the state-of-the-art film growth technique (such as molecular beam epitaxy (MBE) and pulsed laser deposition (PLD)) and the mechanical exfoliation technique in 2D materials have witnessed atomically flat crystalline superconducting films and nanodevices [33-39]. Additionally, the application of new experimental technique enables the measurements in lower temperature regime and higher magnetic fields. Benefit from these developments, unprecedented quantum phenomena have been revealed in the quantum phase transitions. In this review, we focus on recent experimental progress and physical understandings on the quantum phase transitions in 2D superconductors, especially the QGS and the anomalous metal state.

### 1.1 Superconductor-insulator and superconductor-metal transitions

How does the disorder and fluctuation affect the superconducting transition? Anderson proposed that nonmagnetic impurities have no significant effect on the $s$-wave superconductivity in a weakly disordered superconducting system [40]. Increasing the amounts of disorder would result in the localization of electronic wavefunctions [41], and as a result the superconductivity gradually disappears as the disorder increases [42-44]. In a 2D system, the Hohenberg-Mermin-Wagner theorem suggests the fluctuation is so strong that the long-range order cannot exist [45, 46] if the spin-orbit coupling and anisotropy are not considered.

However, in 2D superconducting systems, superconductivity with quasi-long-range order could survive due to the topological BKT transition [21-23].

Experimentally, the effect of disorder and fluctuation on superconductivity has been widely investigated since 1980s. In amorphous superconducting thin film, it was found that the transition temperature $T_c$ falls with increasing disorder (the disorder is characterized by the normal state sheet resistance) [47]. In 1989, Haviland, Liu, and Goldman reported the experimental evidence that the disorder and fluctuation could be strong enough to induce the SIT at low temperatures in amorphous bismuth films by studying the temperature dependent resistance $R(T)$ with different film thicknesses [1]. As shown in Fig. 1(a), SIT occurs as film thickness is reduced. The critical resistance of SIT in this report is close to the quantum resistance of Cooper pairs $\frac{h}{(2e)^2}$, originating from the duality between Cooper pairs and vortices [48].

The discovery of SIT paves the way for studying quantum phase transitions in superconducting systems, and thus inspires tremendous interests in both experimentalists and theorists. Yazdani and Kapitulnik observed the magnetic field induced quantum phase transition in amorphous MoGe (α-MoGe) films, where the critical resistance is much smaller than the quantum resistance [4]. The phrase SMT is used to describe such transition from a superconducting state to a weakly localized metallic state, where this metallic state means $k_F l \gg 1$ (where $k_F$ is the Fermi momentum and $l$ is the mean free path in the normal state) with low critical resistance $R_c < \frac{h}{(2e)^2}$ [14]. During the SMT, Cooper pairs break into fermionic quasiparticles and the duality between Cooper pairs and vortices is absent [49]. Thus, the critical resistance of SMT is no longer a universal constant of $\frac{h}{(2e)^2}$ since fermionic channels contribute to the conductance of the system.

The quantum phase transition from a superconducting state to an insulating or metallic state could be induced by tuning various external parameters such as the amount of disorder, the magnetic field, and carrier density (Fig. 1). Specifically, disorder is stronger and quantum fluctuation has a more significant effect in thinner films. Consequently, quantum phase transition could be driven by varying the film thickness. Figure 1(a) presents the SIT in amorphous Bi films with decreasing film thickness from 74.27 Å to 4.36 Å [1]. Another technique to induce the disorder in superconductors is impurity doping. For example, SIT was also observed in $MgB_2$ films by doping carbon impurities (Fig. 1(b)) [50]. Furthermore, high magnetic field was also widely utilized to modulate the quantum phase transitions in 2D superconducting systems. Figure 1(c) shows a typical example of magnetic field induced SMT in α-MoGe film [4]. Besides, quantum phase transition could be driven by tuning the carrier density of the system. For instance, SIT was realized by applying ion gate voltage in ZrNCl nanoflakes (Fig. 1(d)) [38] as well as post annealing in monolayer $Bi_2Sr_2CaCu_2O_{8+\delta}$ (BSCCO) (Fig. 1(e)) [39].

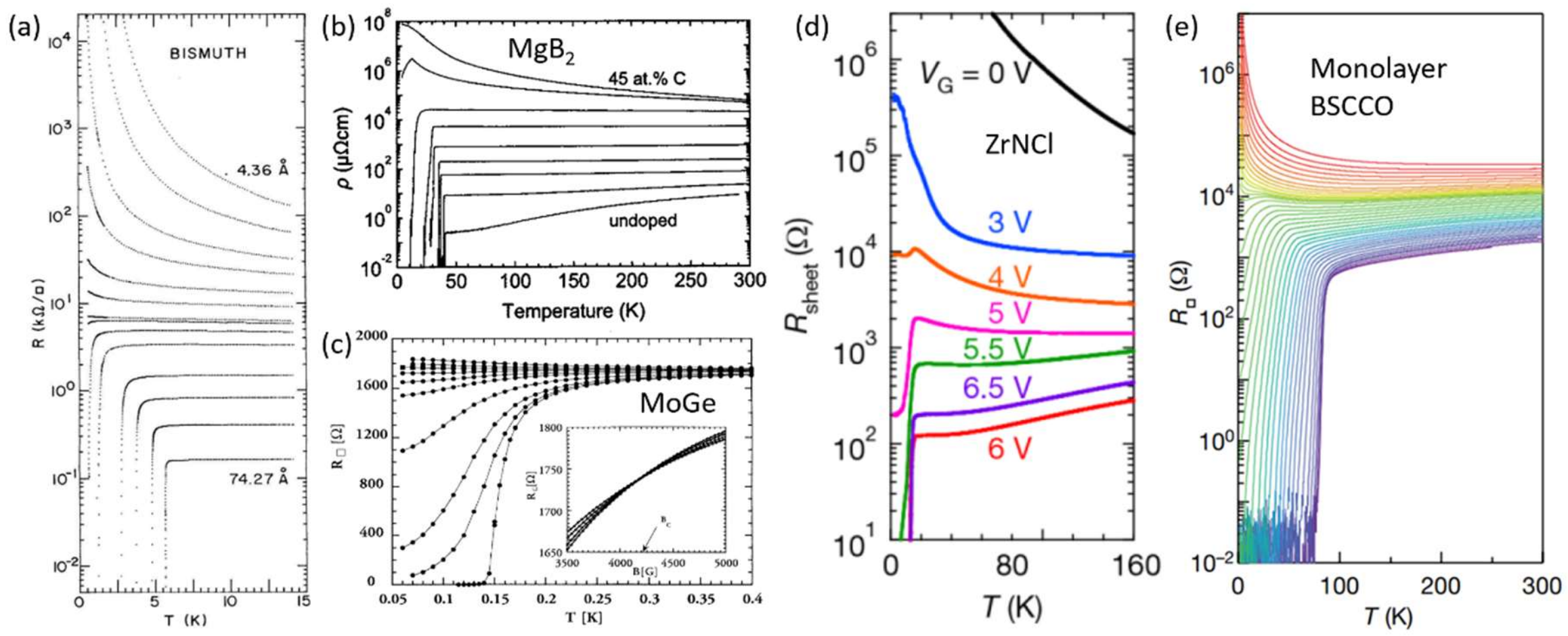

**Figure 1.** Quantum phase transition from a superconducting state to an insulating or metallic state achieved by tuning external parameters. (a) SIT induced by decreasing thickness in amorphous Bi films. Film thicknesses range from 4.36 Å to 74.27 Å. Reprinted from [1]. (b) SIT induced by carbon doping in $MgB_2$ films. Reprinted from [50]. (c) Magnetic field driven SMT in α-MoGe film. Reprinted from [4]. (d) Ion gate voltage tuned SIT in the ZrNCl nanoflake. Reprinted from [38]. (e) SIT achieved by post annealing in the monolayer BSCCO. Reprinted from [39].

## 1.2 Quantum phase transition and quantum criticality

Generally, classical phase transitions can be classified into first-order and continuous transitions (*i.e.*, second-order and higher-order transitions). The first-order transition (*e.g.*, the transition between water and ice) is featured as non-zero latent heat and coexistence of two phases at the transition temperature. On the contrary, latent heat is absent and two phases cannot coexist in the continuous transition. Typical examples of continuous transitions include superconducting transition[1] and ferromagnetic transition by varying temperature in the absence of magnetic field. Landau theory has made a great success in predicting the mean-field behavior of continuous phase transition by introducing the order parameter [51]. The phase transition occurs when thermal fluctuations destroy the long-range order and the amplitude of order parameter drops to zero. In continuous phase transition, the system undergoes the transition continuously from an ordered state (with non-zero order parameter) to a disordered state (with zero order parameter).

Distinct from the classical phase transition, the quantum phase transition takes place at zero Kelvin, when the ground state changes in response to a control parameter [52-57]. The continuous quantum phase transition from an ordered ground state to a quantum disordered state (Fig. 2(a)) can be described in the standard Landau-Ginzburg-Wilson (LGW) framework [58], which links SIT/SMT to other systems exhibiting continuous quantum phase transitions, such as superfluid helium, various magnetic systems that cross between different magnetic ground states [59-61], and $ABO_3$-type perovskites that undergo the ferroelectric to paraelectric transition [62, 63]. Quantum phase transitions could also occur in systems involving topological order, such as 2D electron systems that undergo the transition between quantized Hall plateaus at high magnetic fields and ultralow temperatures [64].

---

[1] To be specific, the transition from superconducting state to normal state in the absence of magnetic field is second order transition. For type-I superconductors, the transition in magnetic field is of first order; for type-II superconductors, the transition in magnetic field is of second order.

In the following, we briefly introduce the critical behavior near the critical point of a continuous phase transition. As the critical point is approached, the spatial correlation length $\xi$ and the temporal correlation length $\xi_\tau$ diverge as [55, 56]

$$\xi \propto |\delta|^{-\nu}, \xi_\tau \propto \xi^z \propto |\delta|^{-z\nu}. \qquad (1)$$

Here, $\delta = |p - p_c|/p_c$ is the dimensionless parameter with the critical parameter $p_c$, $\nu$ is the correlation length critical exponent, and $z$ is the dynamical critical exponent. To determine whether the quantum effect is important to the critical behavior, it is useful to compare two energy scales, namely the energy scale of quantum fluctuation $\hbar\omega_c = h/\xi_\tau \propto h|\delta|^{z\nu}$ and the energy scale of thermal fluctuation $k_B T$. Thus, for any phase transition occurring at finite temperature ($T_c > 0$), the classical thermal fluctuation is dominant when the system is sufficiently close to the critical point ($|\delta| < (k_B T_c/h)^{1/z\nu}$). On the other hand, for the phase transition occurring at zero Kelvin, the critical behavior is always dominated by the quantum fluctuation. This justifies the reason why the zero-temperature phase transition is called the quantum phase transition. Although zero Kelvin is never reachable in real systems, the emergent excitations of quantum criticality can profoundly influence the physical properties at finite temperatures in quantum critical region [65-67]. The boundary of the quantum critical region is determined by $\hbar\omega_c \sim k_B T$ [56] (dashed black curves in Fig. 2(a)).

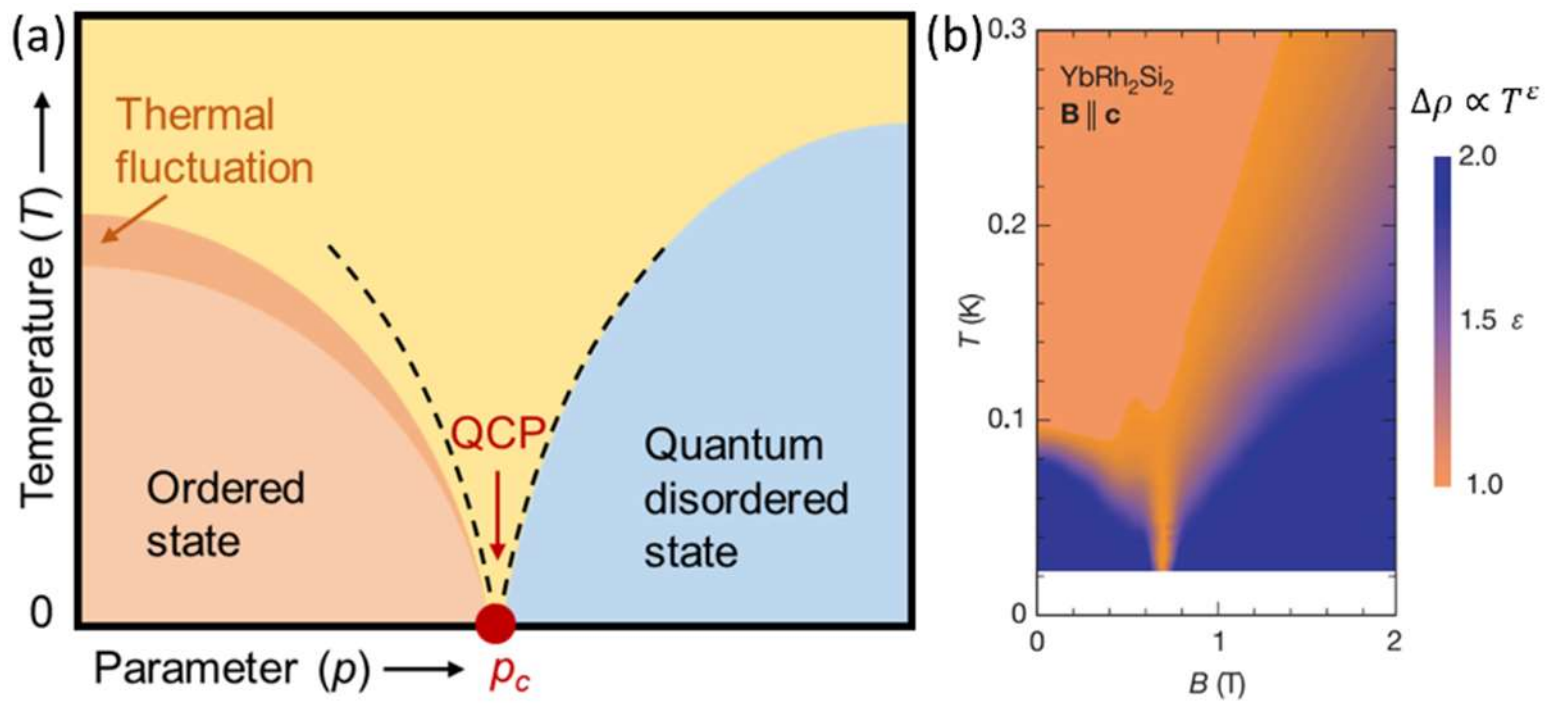

**Figure 2.** (a) Schematic for continuous quantum phase transition. The quantum phase transition occurs at zero Kelvin, when the control parameter is tuned across the quantum critical point $p_c$. (b) Antiferromagnetic to paramagnetic phase transition is a typical example of continuous quantum phase transition from an ordered ground state to a quantum disordered state. Temperature-field phase diagram of heavy fermion material $YbRh_2Si_2$ shows a singular

quantum critical point which gives rise to the strange metal behavior in the quantum critical regime. Figure (b) is reprinted from [60].

In recent years, the experimental indications of quantum criticality have been demonstrated in various condensed matter systems [59-63, 68-72]. For instance, in heavy fermion material $YbRh_2Si_2$, a low magnetic field could drive a quantum phase transition from an antiferromagnetic phase to a paramagnetic phase [60]. Strange metal behavior with linear-in-temperature resistance ($\Delta\rho \propto T^{\varepsilon}, \varepsilon = 1$) appears in the quantum critical region (orange area in Fig. 2(b)), which is attributed to the emergent excitations of quantum criticality. In cuprates and iron pnictide superconductors, the strange metal behavior is also detected in the quantum critical region with the critical point lying beneath the superconducting dome [68-71]. Thus, the quantum criticality in diverse systems may give rise to similar properties and universal scaling dependence. Furthermore, Chapline, Laughlin and Santiago proposed that quantum gravity at the horizon of a black hole could be analogous to quantum criticality in the condensed matter [73]. Therefore, the study of quantum criticality in condensed matter systems may deepen our understanding of the matter in the cosmos [74].

### 1.3 Scaling analysis of quantum phase transition

The dynamics and statics decouple in a classical system; thus, the classical phase transition can be described using time-independent theories in $d$ dimensions (where $d$ is the spatial dimensionality of the system). In a quantum system, the kinetic and potential parts of the Hamiltonian do not commute due to the Heisenberg uncertainty principle. Consequently, the dynamics and statics are intertwined in the quantum phase transition; both temporal and spatial dimensions need to be considered in the quantum phase transition. Note that the canonical density operator $e^{-H/k_BT}$ is the same as the time evolution operator $e^{-iH\tau/\hbar}$ if one identifies imaginary time $\tau = -i\hbar/k_BT$ [56]. A $d$-dimensional quantum system could thus be treated as a classical system of $d + z$ dimensions, where the temporal dimension $z$ comes from the power-law relation between the temporal correlation length and the spatial correlation length ($\xi_\tau \propto \xi^z$), and the size of the imaginary time is given as $L_\tau = \hbar/k_B T$.

In the investigation of SIT/SMT, the scaling dependence of resistance is developed in terms of both the correlation length critical exponent $\nu$ and the dynamical critical exponent $z$ [2-4]:

$$R = R_c F(\frac{\delta}{T^{1/z\nu}}, \frac{\delta}{E^{1/(z+1)\nu}}). \qquad (2)$$

Here, $\delta = |p - p_c|/p_c$ is the dimensionless parameter with the critical parameter $p_c$, $R_c$ is the critical resistance of the transition, $F$ is the scaling function, $T$ and $E$ are temperature and electric field of the measurement. Since the real material cannot be cooled down to zero Kelvin, the temporal dimension at finite temperature is of finite size. It is thus known as the finite size scaling.

Experimentally, Hebard and Paalanen reported the scaling dependence of resistance in SIT (Fig. 3) [2]. A crossing point of magnetoresistance curves at various temperatures was identified as the quantum critical point in amorphous $InO_x$ films. Scaling analysis of magnetoresistance curves at various temperatures was shown to collapse on a single function, which is consistent with equation (2) and thus provides information on the critical exponent $z\nu$. A great amount of work has been accomplished to confirm the scaling behavior of SIT/SMT in diverse superconducting systems [2, 4]. The early studies on disordered superconducting films tend to show a single crossing point in the magnetoresistance curves. However, double crossing points were reported in superconducting oxide interfaces [12] and copper oxide superconducting films [13]. More interestingly, a continuous line of crossing points was observed in magnetic field induced SMT of atomically thin Ga films [14] in 2015, which leads to the discovery of QGS of SMT in 2D crystalline superconducting films (the focus of Chapter 2).

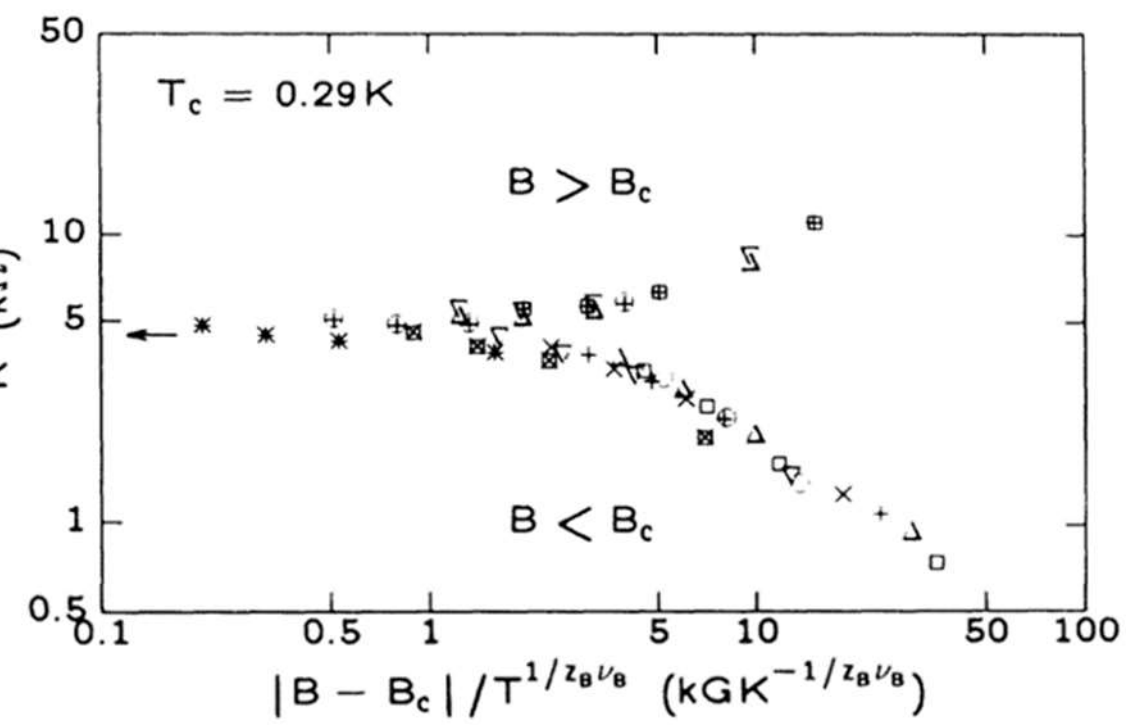

**Figure 3.** Scaling dependence of resistance in SIT of amorphous $InO_x$ film. Reprinted from [2].

## 2. Quantum Griffiths singularity

Randomness and disorder are essential elements in real condensed matter systems and play prominent roles in phase transitions [75-78]. Particularly, rare regions are found to significantly influence the classical phase transitions and lead to singular free energy (*i.e.*, the Griffiths singularity). In quantum phase transitions, the rare regions could give rise to QGS characterized by a divergent dynamical critical exponent approaching the infinite-randomness quantum critical point. In this chapter, we will briefly review representative theoretical developments related to Griffiths singularity and QGS (Chapter 2.1), followed by experimental indications of QGS in ferromagnetic quantum phase transitions (Chapter 2.2). In Chapter 2.3, we will focus on experimental discovery and progress of QGS in SMT of 2D superconducting systems.

### 2.1 Theoretical developments of quantum Griffiths singularity

#### 2.1.1 The effect of rare regions on classical and quantum phase transition

Harris proposed a criterion to determine the stability of the critical point against the quenched disorder [79]. If the correlation length exponent $\nu$ and the dimensionality $d$ fulfill the Harris criterion $d\nu > 2$, the critical behavior is basically unaffected by the disorder. On the contrary, if the critical region violates the Harris criterion ($d\nu < 2$), the critical behavior is destabilized by the disorder[2]. Harris criterion was originally proposed in classical phase transitions, but it has the same form in quantum phase transitions since in most cases the disorder in condensed matter systems is time-independent over the measurement timescales [76].

In 1969, in the study of randomly diluted Ising magnet, Griffiths pointed out that the rare region could lead to singular free energy in classical phase transition [80], which is called as the Griffiths singularity. The presence of random vacancies reduces the ferromagnetic transition temperature from $T_M^0$ to $T_M$. In the intermediate temperature ($T_M < T < T_M^0$), although the system is globally in the paramagnetic phase, there exist locally ferromagnetic rare regions. Such rare regions have slower dynamics compared to the global system and hence give rise to singular free energy. The effect of rare region was also noticed by McCoy and Wu in a 2D classical Ising model with linear defects [81]. The Griffiths effect is thus sometimes called as the Griffiths-McCoy-Wu effect.

The rare regions have stronger effect in quantum phase transitions considering the infinite size of the imaginary time direction approaching zero Kelvin. In 1992, Fisher developed a real-space renormalization group theory to understand the critical behavior in random transverse-field Ising model [82, 83], which gives rise to infinite-randomness quantum critical point in the strong randomness regime. In the vicinity of such quantum critical point, the divergence of the temporal correlation length $\xi_\tau$ follows an activated scaling $ln(\xi_\tau) \propto \xi^\psi$ (where $\xi$ is the spatial correlation length and $\psi$ is the tunneling exponent) rather than the conventional power-law scaling $\xi_\tau \propto \xi^z$ (where $z$ is the dynamical critical exponent) [82-84]. Noteworthily, the activated scaling indicates an exponentially slow dynamics of spin flipping at the quantum critical point for a large spin cluster with length scale $\xi$.

Vojta and Hoyos proposed a unified classification of quantum phase transitions in disordered systems [85]. Particularly, if Harris criterion is violated, the activated scaling $ln(\xi_\tau) \propto \xi^\psi$ gives rise to divergent dynamical critical exponent $z \propto |\delta|^{-\nu\psi}$ as the system approaches the infinite-randomness quantum critical point (where $\delta = |p - p_c|/p_c$ is the dimensionless parameter with the critical parameter $p_c$) [85]. The divergence of dynamical critical exponent has been identified as the primary feature of QGS.

[2] If the Harris criterion is violated, the clean correlation length exponent $\nu$ would shift to a new value $\nu'$ in the presence of disorder under renormalization group.

The critical exponents $\alpha, \beta, \gamma, \delta, \eta, \nu, z$ are utilized to characterize the universality class of phase transitions. Due to the scaling relations, commonly three of these parameters are independent. For the case of QGS, the system manifests the activated scaling behavior and the conventional scaling law breaks down [83, 86]. The independent critical exponents for this activated scaling are $\nu$, $\psi$ and $\phi$, which correspond to spatial correlation length, temporal correlation length and rare cluster volume, respectively. The critical exponents $\alpha$, $\gamma$, and $\delta$ are not suitable for quantum Griffiths phase, with the value changing when approaching the infinite randomness critical point. On the other hand, the values of $\eta$ and $\beta$ can be obtained based on the hyperscaling relations $\eta = 2(d - \psi \cdot \phi)$ and $\beta = \nu(d - \psi \cdot \phi)$ [86], where $d$ is the dimensionality. The critical exponents for the QGS in 1D, 2D and 3D random transverse-field Ising model are summarized in Table 1.

**Table 1.** The critical exponents for the quantum Griffiths singularity in random transverse-field Ising model for the systems with different dimensionalities. The exponents in 1D systems have analytic results [82, 83] and those in 2D and 3D systems were calculated using numerical renormalization group method [84, 87, 88].

| Exponents | Definition | Theoretical value |
|---|---|---|
| $\nu$ | Spatial correlation length $\xi \propto |\delta|^{-\nu}$ | 1D: 2<br>2D: 1.24<br>3D: 0.97 |
| $\psi$ | Temporal correlation length $ln(\xi_\tau) \propto \xi^{\psi}$ | 1D: 0.5<br>2D: 0.48<br>3D: 0.46 |
| $\phi$ | Rare cluster volume $m \propto [ln(\xi_\tau)]^\phi$ | 1D: $\frac{1+\sqrt{5}}{2}$<br>2D: 2.12<br>3D: 2.52 |
| $\eta$ | Correlation function $G(r) \propto r^{-\eta}$ | 1D: $\frac{3-\sqrt{5}}{2}$<br>2D: 1.96<br>3D: 3.68 |
| $\beta$ | Magnetization $m \propto |\delta|^\beta$ | 1D: $\frac{3-\sqrt{5}}{2}$<br>2D: 1.22<br>3D: 1.78 |

The rare regions in quantum Griffiths phase could also result in unconventional thermodynamic properties close to the quantum critical point [76, 89]. The probability of finding a rare region $P$ decreases exponentially with the volume of rare region $V$, $P \propto \exp(-bV)$, where $b$ is a parameter that characterizes the strength of disorder. The characteristic energy scale of rare region $E$ also exponentially depends on its volume $V$, $E \propto \exp(-cV)$. These two exponential relations yield an energy spectrum $P(E) \propto E^{b/c-1} = E^{\lambda-1}$ with the Griffiths exponent $\lambda = b/c = d/z$ (where $d$ is the spatial dimensionality and $z$ is the dynamical critical exponent) [90]. The energy spectrum leads to power-law behaviors of specific heat $C_V \propto T^{\frac{d}{z}}$, magnetization $M \propto H^{\frac{d}{z}}$ and susceptibility $\chi \propto T^{\frac{d}{z}-1}$ with $z \propto |\delta|^{-\nu}$ [76].

### 2.1.2 Quantum Griffiths singularity in superconductor-metal transitions

Early theoretical investigations revealing the influence of quenched disorder on the quantum phase transition and the QGS are mainly based on the Ising model with discrete symmetry. Table 2 summarizes representative theoretical models for the QGS [82-88, 91-99]. Particularly, Vojta, Kotabage and Hoyos developed a strong-disorder renormalization group method to study the quantum phase transitions with continuous $O(N)$ symmetry order parameters including the SMT (Table 2) [96, 97]. This work and subsequent numerical studies [98, 99] reveal the possibility to study QGS in superconducting systems from the theoretical perspective.

**Table 2.** Representative theoretical models for the quantum Griffiths singularity.

| **Theoretical models** | **Description of the model** | **Existence of quantum phase transition and quantum Griffiths singularity** |
|---|---|---|
| **Random transverse-field Ising model** | The Hamiltonian reads $H = -\sum_{(ij)} J_{ij}\sigma_i^z\sigma_j^z - \sum_i h_i\sigma_i^x$, where the sum $\Sigma_{(ij)}$ is over the nearest sites, $\sigma_i$ are Pauli matrices, $J_{ij}$ and $h_i$ are random interactions and random transverse fields. The quenched disorder effect is induced by $J_{ij}$ and $h_i$. | The quantum phase transition in *d*-dimensional system is equivalent to the classical phase transition in a (*d*+1)-dimensional classical Ising model. And the *d*-dimensional random transverse-filed Ising model is reminiscent of the (*d*+1)-dimensional McCoy-Wu model with correlated disorder. Quantum Griffiths singularity exists when approaching the infinite randomness critical point of the quantum phase transition.<br><br>Representative theoretical studies:<br>1D systems: Real-space renormalization group method [82, 83]; Numerical study [91]; See also the review paper [86].<br>2D systems: Numerical renormalization group method [84, 87]; Quantum Monte Carlo [92].<br>3D systems: Numerical renormalization group method [88]. |
| **Correlated disordered XY model** | The Hamiltonian reads $H = -\sum_i(\frac{1+\gamma_i}{2}\sigma_i^x\sigma_{i+1}^x + \frac{1-\gamma_i}{2}\sigma_i^y\sigma_{i+1}^y + h_i\sigma_i^z)$, where $\sigma_i$ are Pauli matrices, $\gamma_i$ and $h_i$ are random interactions and random fields. The quenched disorder effect is induced by $\gamma_i$ and $h_i$. | In the presence of quenched disorder, the ferromagnetic quantum phase transition is in the same universality class as the random transverse-field Ising model. The quantum Griffiths singularity exists near the quantum critical point of ferromagnetic quantum phase transition.<br><br>Representative theoretical studies:<br>Analytic studies [93, 94]; Fidelity approach [95]. |
| **Heisenberg model with Ohmic dissipation (more generally, continuous symmetry model)** | The model considers *N*-component vector order parameter $\varphi = (\varphi_1, \ldots, \varphi_N)$ in $d$ space dimensions. This model applies to superconductor-metal transition for $N = 2$ and antiferromagnetic quantum phase transition for $N = 3$. The action of the Landau-Ginzburg-Wilson field theory reads [97] $S = \int dydx\varphi(x)\Gamma(x,y)\varphi(y) + \frac{u}{2N}\int dx\varphi^4(x)$, where $x = (\boldsymbol{x}, \tau)$ comprises position $\boldsymbol{x}$ and imaginary time $\tau$. The Fourier transform of the bare inverse propagator $\Gamma(x,y)$ is $\Gamma(\boldsymbol{q}, \omega_n) = r + \xi_0^2\boldsymbol{q}^2 + \gamma_0\lvert\omega_n\rvert$. In the presence of quenched disorder, the distance from criticality *r*, the microscopic length scale $\xi_0$, the Ohmic dissipation damping coefficient $\gamma_0$ and the standard quartic coefficient $u$ are random functions of spatial position. | The quantum phase transition in this model is in the same universality class as the random transverse-field Ising model in the same space dimensionality. This model displays the quantum Griffiths singularity in the presence of Ohmic dissipation.<br><br>Representative theoretical studies:<br>Strong disorder renormalization group method [96, 97]; Numerical studies [98, 99]; See also Chapter 21.4 in Sachdev "Quantum Phase Transitions" [54]. |

In a 2D superconducting system, at magnetic fields slightly above the mean-field upper critical field ( $B_{c2} < B < B_c^*$ ), although the long-range superconducting order is destroyed, superconducting rare regions emerges with phase coherence inside the region (see Fig. 4(a) for a schematic of superconducting rare regions). Large superconducting rare regions have slower dynamics compared to the global system and hence dramatically influence the dynamical critical behavior in the quantum phase transition. In the study of SMT in atomically thin crystalline Ga films, it is found that the phase boundary deviates significantly from the mean-field behavior (see Fig. 4(b) for a schematic of the phase diagram of SMT in Ga films). More interestingly, the dynamical critical exponent is found to diverge when approaching the quantum critical point [14], which marks the experimental discovery of QGS of SMT in 2D superconducting systems.

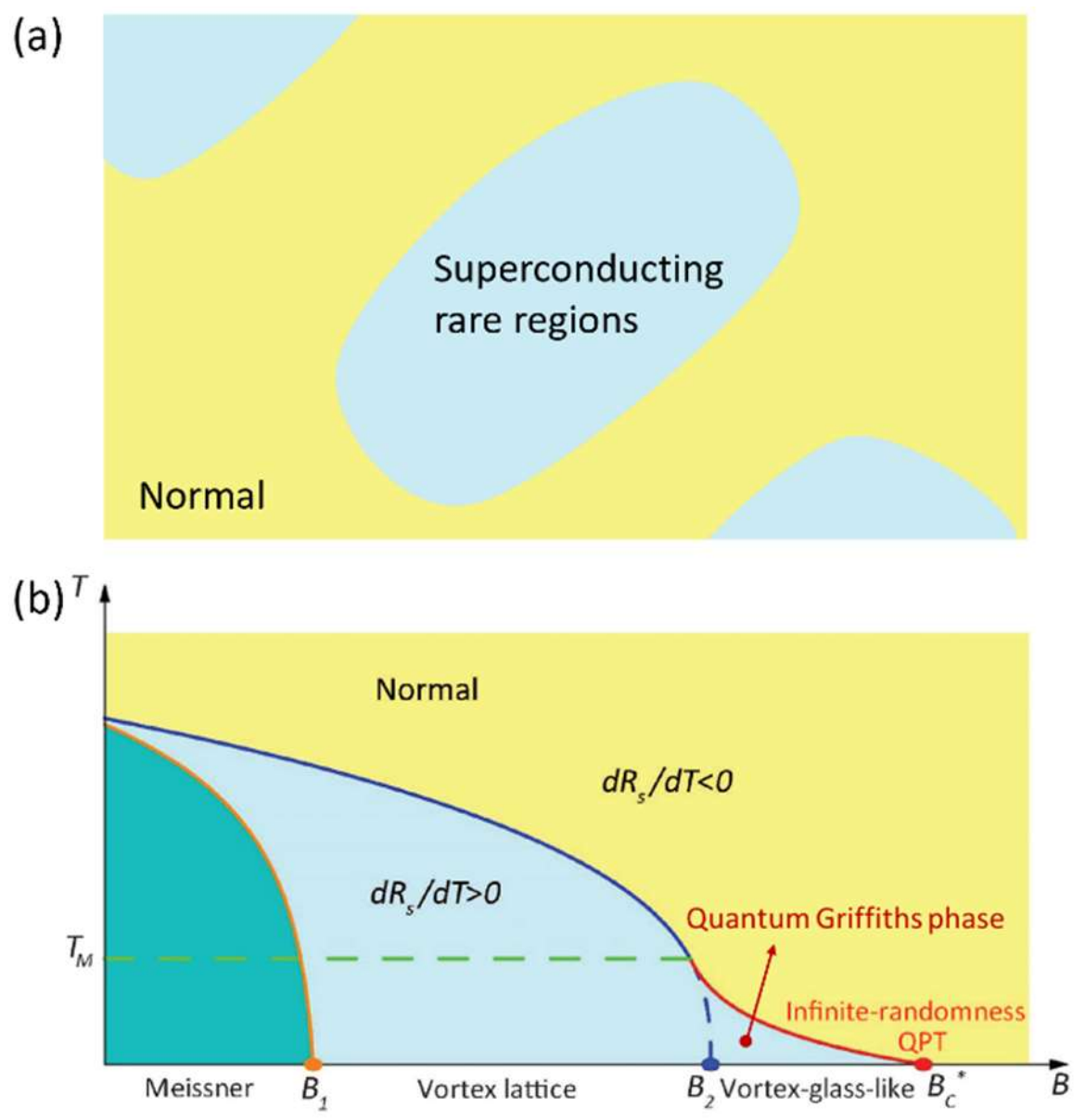

**Figure 4.** (a) A schematic of superconducting rare regions in 2D systems. (b) In SMT of 2D system, the superconducting rare regions have a significant influence on the boundary of phase diagram and give rise to the activated scaling behavior when approaching the infinite-randomness quantum critical point $B_c^*$. Figure (b) is reprinted from [14].

## 2.2 Experimental signatures of quantum Griffiths phase in ferromagnetic quantum phase transitions

Experimental indications of quantum Griffiths phase and power-law QGS of the Griffiths exponent have been reported in ferromagnetic quantum phase transitions [90, 100-104]. Ferromagnetic quantum phase transition was realized in $CePd_{1-x}Rh_x$ alloy by tuning Rh concentration $x$. By performing measurements of resistivity, susceptibility, specific heat and thermal expansion, the ferromagnetic phase boundary was found to deviate from the conventional behavior and develop a long tail towards large $x$ at low temperatures [103]. In the tail region, the Grüneisen parameter (defined as the ratio between thermal expansion coefficient and specific heat) depends logarithmically on temperature [104], which is in agreement with the critical behavior predicted in quantum Griffiths phase [89]. Stronger indication of QGS was demonstrated in binary alloy $Ni_{1-x}V_x$ [90, 100]. The doping of vanadium could introduce strong disorder and induce a ferromagnetic to paramagnetic quantum phase transition. Exponents $\alpha$ or $(1-\gamma)$, *i.e.* the Griffiths exponent $\lambda$, was determined from the magnetization $M(H)$ curve. To be specific, $\alpha$ is defined by $M \propto H^{\alpha}$ on the paramagnetic side or $M - M_0 \propto H^{\alpha}$ on the ferromagnetic side [100], and $\gamma$ is defined by $M/H \propto T^{-\gamma}$ [90]. The $x$ dependence of exponents $\alpha$ and $1-\gamma$ were found to follow the power-law relation $\alpha(x) \propto |x - x_c|^{\nu\psi}$ and $1-\gamma(x) \propto |x - x_c|^{\nu\psi}$ (Fig. 5), indicating the power-law QGS close to the quantum critical point $x_c$ in $Ni_{1-x}V_x$ alloy.

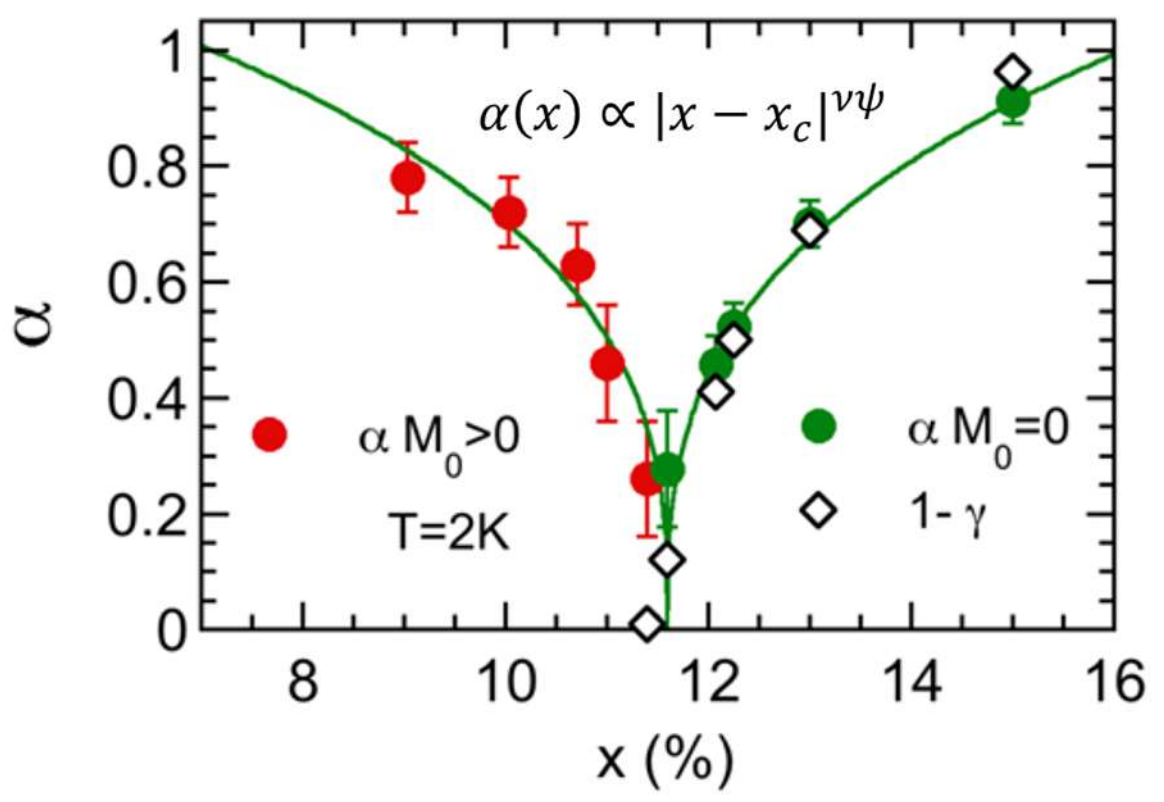

**Figure 5.** Power-law QGS close to the quantum critical point $x_c$ in $Ni_{1-x}V_x$ alloy. The $x$ dependence of exponent $\alpha$ and $1-\gamma$ follow the power-law relation $\alpha(x) \propto |x - x_c|^{\nu\psi}$ and $1-\gamma(x) \propto |x - x_c|^{\nu\psi}$, consistent with quantum Griffiths model. $\alpha$ is defined by $M \propto H^{\alpha}$ on the paramagnetic side or $M - M_0 \propto H^{\alpha}$ on the ferromagnetic side [100], and $\gamma$ is defined by $M/H \propto T^{-\gamma}$ [90]. Reprinted from [100].

### 2.3 Quantum Griffiths singularity in 2D superconductors

#### 2.3.1 Experimental discovery of quantum Griffiths singularity in crystalline 2D Ga superconducting films

We have demonstrated in Chapter 2.1 that the rare regions could dramatically alter the critical behavior of quantum phase transition and give rise to QGS. Experimental indications, such as the power-law QGS of the Griffiths exponent ($\lambda \propto |\delta|^{\nu\psi}$), have been reported in 3D magnetic systems [90, 100]. However, the primary feature of QGS, namely the divergence of dynamical critical exponent, has not yet been observed in such systems. The 2D superconductors provide another promising platform to study quantum phase transitions due to the strong fluctuations in reduced dimensions. In 2015, Xing *et al.* reported the discovery of QGS of SMT in ultrathin crystalline Ga films [14, 105] by experimentally observing the divergence of dynamical critical exponent. In this work, magnetic field is used to drive the quantum phase transition continuously across the quantum critical point.

Figure 6 presents the magnetoresistance of three-monolayer (ML) thick Ga film grown on GaN(0001) substrate by MBE [14]. Distinct from the previous reports [2, 4, 12, 13], the magnetoresistance curves at various temperatures ranging from 0.025 K to 2.8 K exhibit a series of crossing points that form a continuous line of SMT "critical" points (Fig. 6). The inset of Fig. 6 shows the temperature dependence of the "critical field" $B_c$ for the crossing points. $B_c$ follows the trend given by the mean-field Werthamer-Helfand-Hohenberg (WHH) theory [106] above 1 K, but turns upward and deviates significantly from the WHH formula at lower temperatures, suggesting the formation of quantum Griffiths phase in the ultralow temperature regime.

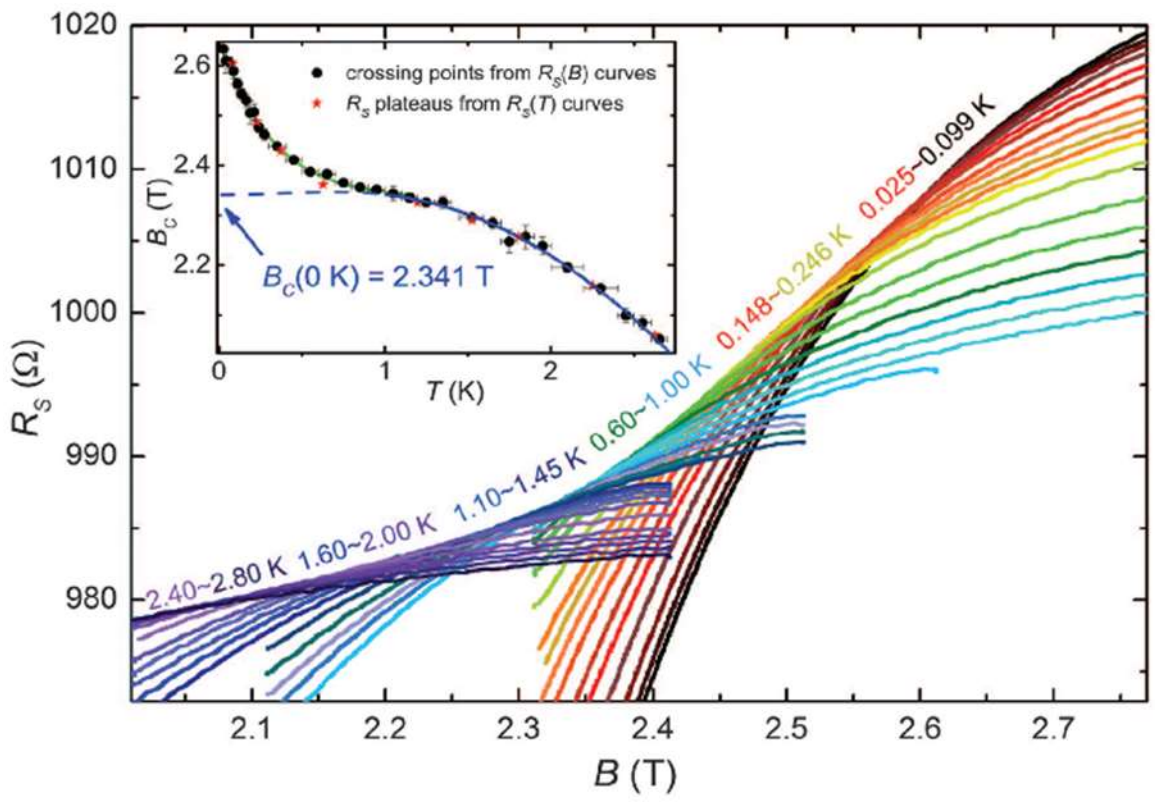

**Figure 6.** Magnetic field dependence of sheet resistance at various temperatures from 0.025 K to 2.80 K in 3-ML Ga film, exhibiting a continuous line of SMT "critical" points. Inset, the temperature ($T$) dependence of the "critical field" ($B_c$) for the crossing points. Reprinted from [14].

Experimental evidence of QGS comes from the scaling analysis of the critical exponent $z\nu$. Adjacent magnetoresistance curves are considered as one group, which shows approximately one crossing point ($B_c$, $R_c$). Magnetoresistance curves around ($B_c$, $R_c$) follow the finite size scaling

$$\frac{R(B-B_c,T)}{R_c} = F\left(\frac{B-B_c}{T^{1/z\nu}}\right). \qquad (3)$$

Here, $F$ is an arbitrary function with $F(0) = 1$, and the scaling analysis yields the critical exponent $z\nu$ at $B_c$. Nine representative crossing points on the phase boundary are selected, and the relation between $z\nu$ and magnetic field is obtained by finite size scaling analysis (Fig. 7). The divergence of the critical exponent is consistent with the random transverse-field Ising model approaching the infinite-randomness quantum critical point [84, 97]

$$z\nu \propto |B - B_c^*|^{-\nu\psi}. \qquad (4)$$

In a 2D system, the model yields $\nu \approx 1.2$ and $\psi \approx 0.5$ [84, 85]. With these values, the divergent $z\nu$ approaching $B_c^*$ can be well fitted using the model when $z\nu$ is larger than 1 (Fig. 7). The good agreement between experimental observation and theoretical expectation provides strong evidence of QGS of SMT in a 2D superconducting system.

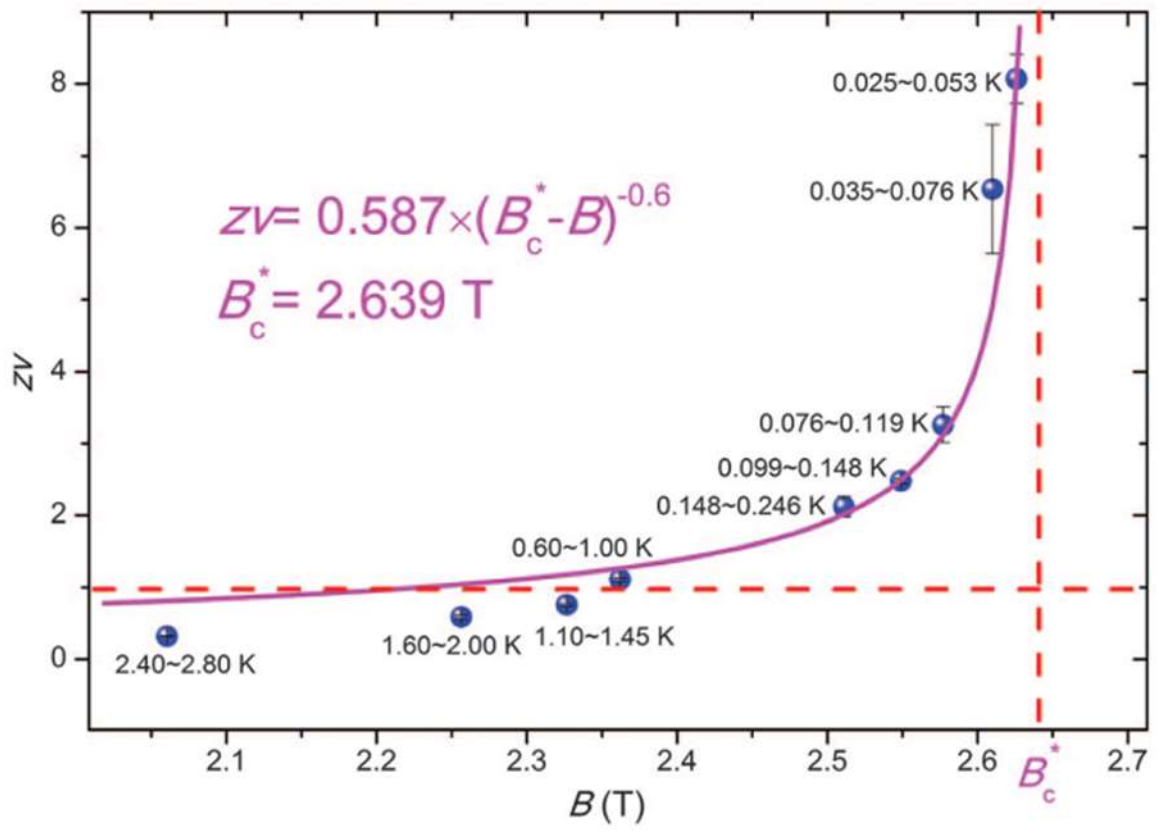

**Figure 7.** Critical exponent $z\nu$ as a function of magnetic field diverges when approaching the quantum critical point $B_c^*$. Reprinted from [14].

### 2.3.2 The universality of quantum Griffiths singularity in 2D superconductors

The discovery of QGS in Ga films has stimulated related research in various 2D superconducting systems [33, 107-118]. As shown in Table 3, QGS has been reported in a wide range of superconducting systems, including atomically flat crystalline superconducting films, exfoliated 2D superconducting devices, superconducting oxide interfaces, polycrystalline and amorphous superconducting films. Although these 2D superconductors exhibit large diversity in film smoothness, fabrication method, critical field, and the shape of phase boundary, all these systems show a divergent critical exponent when approaching the quantum critical point. Particularly, the critical behavior $z\nu \propto |B - B_c^*|^{-0.6}$ has been confirmed in diverse 2D superconducting systems, demonstrating the universality of QGS. Recently, the divergent critical exponent approaching zero temperature has also been reported as a signature of QGS in electronic nematic quantum phase transition of iron-based superconductor $FeSe_{0.89}S_{0.11}$ [119].

**Table 3.** Quantum Griffiths singularity in 2D superconducting systems. The critical exponent $z\nu$ in the table is given by finite size scaling $R = R_c F(\delta/T^{1/z\nu})$, where $z\nu \propto |\delta|^{-\nu}$ diverges approaching the quantum critical point as a result of the activated scaling of temporal correlation length $ln(\xi_\tau) \propto \xi^\psi$. Here, $\delta = |B - B_c|/B_c$ is a dimensionless parameter with the critical field $B_c$, $\xi$ is the spatial correlation length, $z$ is the dynamic critical exponent, $\nu$ is the correlation length critical exponent, and $\psi$ is the tunneling exponent.

| System | Description of the system | Divergence of critical exponent | Description of phase boundary | Reference |
|---|---|---|---|---|
| **Ga films** | Atomically flat crystalline film grown by MBE | $z\nu \propto \|B - B_c^*\|^{-0.6}$ with $B_c^* = 2.64$ T | Crossing point field ($B_c$) turns upward and deviates from WHH formula below 1 K | 14 |
| **Pb films** | Atomically flat crystalline film grown by MBE | $z\nu \propto \|B - B_c^*\|^{-0.6}$ with $B_c^* = 2.96$ T | Phase boundary bends down below 2.4 K | 107 |
| **$NbSe_2$ films** | Atomically flat monolayer film grown by MBE | $z\nu \propto \|B - B_c^*\|^{-0.6}$ with $B_c^* = 3.28$ T | $B_c$ turns upward and deviates from a linear trend below 0.5 K | 108 |
| **$PdTe_2$ films** | Atomically flat crystalline film grown by MBE | $z\nu \propto \|B - B_c^*\|^{-0.6}$ with $B_c^* = 0.99$ T (Perpendicular field) $B_c^* = 15.21$ T (Parallel field). Direct activated scaling analysis with the irrelevant correction is also performed in this work. | $B_c$ is in good agreement with the activated scaling theory | 109 |
| **LAO/STO interface** | Superconducting oxide interface grown by PLD | $z\nu \propto \lvert B - B_c^*\rvert^{-0.6}$ with $B_c^* = 0.42$ T | No upward trend observed at low temperatures | 110 |
| **EuO/KTO interface** | Superconducting oxide interface grown by oxide MBE | $z\nu \propto \lvert B - B_c^*\rvert^{-0.6}$ with $B_c^* = 1.72$ T | $B_c$ turns upward below 0.2 K | 111 |
| **TiO films** | Epitaxial oxide film grown by PLD | $z\nu \propto \lvert B - B_c^*\rvert^{-0.6}$ with $B_c^* = 10.16$ T (42 nm sample) | No upward trend observed at low temperatures | 112 |
| **$InO_x$ films** | Amorphous film grown by electron beam evaporation | $(\frac{1}{z\nu})_{\mathrm{eff}} = (\frac{1}{\nu\psi})_{\mathrm{eff}} \frac{1}{\ln{(T_0/T)}}$ with $(\nu\psi)_{\mathrm{eff}} = 0.62$ and $T_0 =$ 1.21 K. Direct activated scaling analysis is also performed in this work. | $B_c$ is in good agreement with the activated scaling theory | 113 |
| **4Ha-$TaSe_2$ nanodevices** | Exfoliated 2D superconductor | $z\nu \propto \lvert B - B_c^*\rvert^{-0.6}$ with $B_c^* = 1.72$ T (4.3 nm sample) Direct activated scaling analysis with the irrelevant correction is also performed in this work. | $B_c$ is in good agreement with the activated scaling theory | 114 |
| **Ion-gated $MoS_2$** | Exfoliated 2D superconductor tuned by ionic liquid gate | $z\nu \propto \lvert B - B_c^*\rvert^{-0.6}$ with $B_c^* = 9.28$ T | $B_c$ turns upward at low temperatures | 115 |
| **Ion-gated ZrNCl** | Exfoliated 2D superconductor tuned by ionic liquid gate | $z\nu \propto \lvert B - B_c^*\rvert^{-0.6}$ with $B_c^* = 5.47$ T | $B_c$ turns upward at low temperatures | 115 |
| ***β*-W films** | Polycrystalline film grown by magnetron sputtering deposition | $z\nu \propto \lvert B - B_c^*\rvert^{-0.6}$ with $B_c^* = 5.6$ T (Perpendicular field) $B_c^* = 29.65$ T (Parallel field). Direct activated scaling analysis is also performed in this work. | Under perpendicular field, $B_c$ turns upward and deviates from WHH formula below 1.2 K. Under parallel field, $B_c$ is in good agreement with the activated scaling theory. | 116 |
| **$(TOA)_xSnSe_2$** | Organic-inorganic hybrid superlattice | $z\nu \propto \lvert B - B_c^*\rvert^{-0.6}$ with $B_c^* = 6.21$ T | $B_c$ turns upward and deviates from a linear trend below 2.5 K | 117 |
| **Pb islands on graphene** | Superconducting islands array on graphene | $z\nu \propto \lvert B - B_c^*\rvert^{-0.6}$ with $B_c^* = 0.38$ T | $B_c$ turns upward and deviates from WHH formula below 1 K | 118 |

### 2.3.3 Superconducting fluctuations and anomalous quantum Griffiths singularity

In the study of SMT in ultrathin crystalline Pb films, Liu *et al.* reported the observation of QGS with anomalous phase boundary [107]. Distinct from the usual SMT with monotonic phase boundary separating the regions of $dR/dT < 0$ and $dR/dT > 0$, sheet resistance $R_s(T)$ curves in 4-ML Pb films exhibit a remarkable reentrant behavior under magnetic field slightly over 3.3 T where $R_s(T)$ firstly decreases with decreasing temperature, reaches the minimum and then rises at lower temperature (Fig. 8(a)). The magnetoresistance isotherms exhibit a "regular triangle" crossing point region (Fig. 8(b)), which is different from the "upside-down triangle" region in other 2D superconducting systems with usual SMT (see magnetoresistance isotherms of 3-ML Ga film in Fig. 6). The SMT phase boundary, determined from crossing points of adjacent $R_s(B)$ curves, shows an anomalous behavior that bends down at low temperatures (Fig. 8(c)).

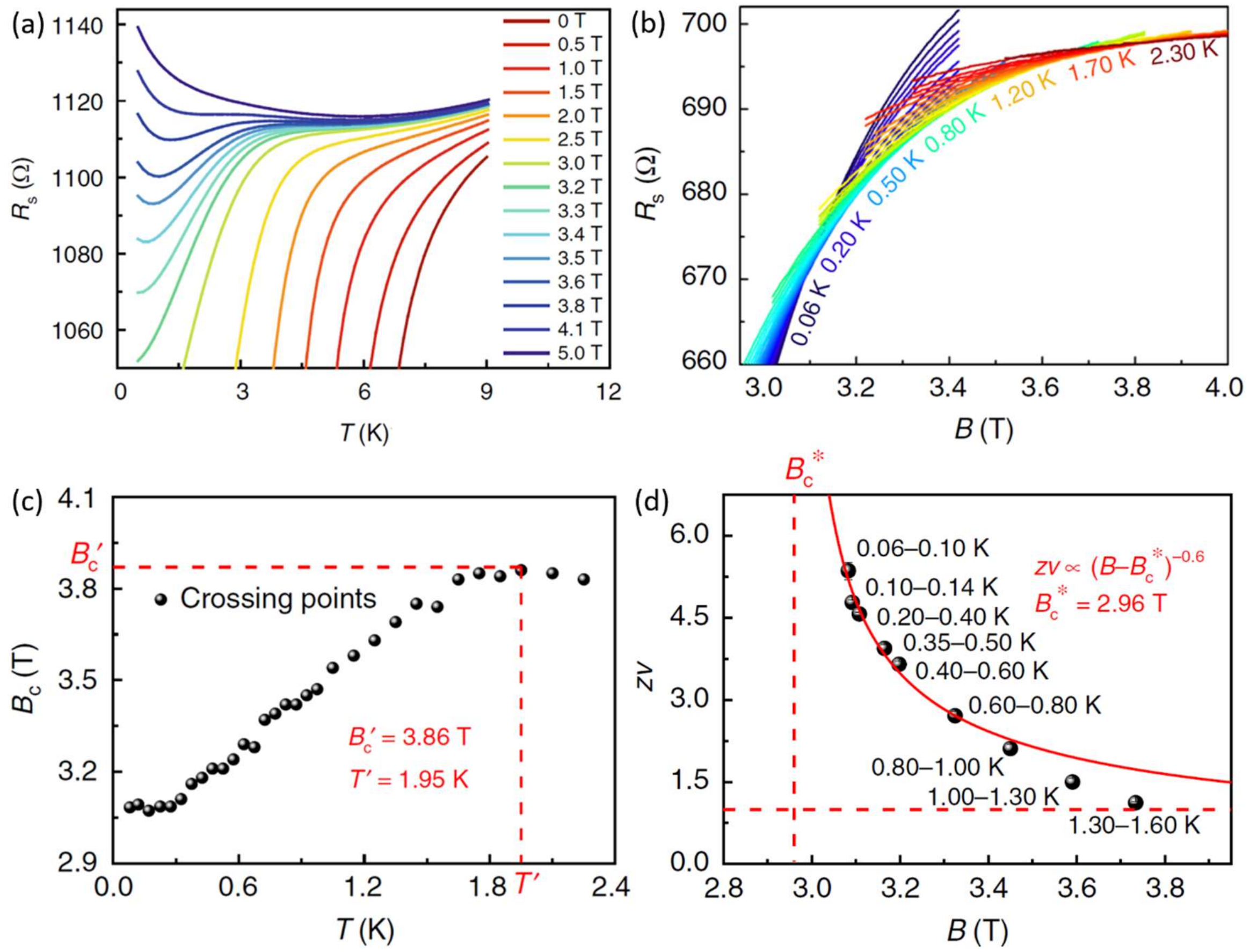

**Figure 8.** QGS with anomalous phase boundary in 4-ML Pb film. (a) Temperature dependence of sheet resistance at various magnetic fields show SMT with reentrant behavior. (b) Magnetic field dependence of sheet resistance at various temperatures from 0.06 K to 2.30 K, showing a "regular triangle" crossing point region. (c) SMT phase boundary determined from crossing points of adjacent $R_s(T)$ curves. The phase boundary shows an anomalous behavior that bends down at low temperature. (d) The critical exponent diverges as $z\nu \propto |B - B_c^*|^{-0.6}$ when approaching the quantum critical point, indicating the anomalous QGS. Reprinted from [107].

For 2D superconductors in the dirty limit[3], the superconducting fluctuation correction to the conductivity is given by Larkin and his collaborators [120, 121]

$$\delta\sigma = \frac{e^2}{\pi^2\hbar}[\alpha I_\alpha(b,t) + \beta I_\beta(b,t)]. \qquad (5)$$

Here, $I_\alpha(b,t) = \ln\frac{r}{b} - \frac{1}{2r} - \psi(r)$, $I_\beta(b,t) = r\psi'(r) - \frac{1}{2r} - 1$, $\alpha$ and $\beta$ are coefficients of superconducting fluctuation term $I_\alpha$ and $I_\beta$, with $r = \frac{b}{3.562t}$, $t = T/T_c \ll 1$, $b = [B - B_{c2}(T)]/B_{c2}(0) \ll 1$, $\psi(r)$ is the digamma function, and $B_{c2}(T)$ is the upper critical field given by WHH theory [106]. The superconducting fluctuation term is related to the Cooperon propagator. Meanwhile, the spin-orbit interaction (SOI) can alter the form of Cooperon propagator, introducing the spin-triplet channel with negative sign in addition to the spin-singlet channel with positive sign [122]. Thus, the pronounced SOI in ultrathin Pb films, evidenced by the observation of Zeeman-protected Ising superconductivity [123], can have a significant influence on $\alpha$ and $\beta$ and hence change the extent of superconducting fluctuation. The anomalous phase boundary as well as the reentrant behavior can be reproduced after taking the superconducting fluctuation and the effect of SOI into account [107]. The enhanced effect of superconducting fluctuation in ultrathin Pb films could be attributed to the relatively low mobility and strong SOI.

---

[3] The criterion for dirty limit in superconductors can be characterized by the ratio between the Pippard coherent length $\xi_0$ and the mean free path *l*. The dirty limit means $\frac{\xi_0}{l} \gg 1$.

Under the influence of superconducting fluctuations with strong SOI, the mean-field phase boundary buckles outward and the anomalous phase boundary exists in the regime between $B_c^*$ and $B_c'$ (Fig. 9). When approaching the infinite-randomness quantum critical point $B_c^*$ along this anomalous phase boundary, the critical exponent diverges as $z\nu \propto |B - B_c^*|^{-0.6}$ (Fig. 8(d)), indicating the anomalous QGS behavior [107]. This experiment demonstrates the universality of the underlying mechanism of QGS (*i.e.*, large superconducting rare regions emerge and dominate the dynamical critical behavior approaching zero Kelvin), even in the 2D superconducting systems with strong fluctuation effect and SOI.

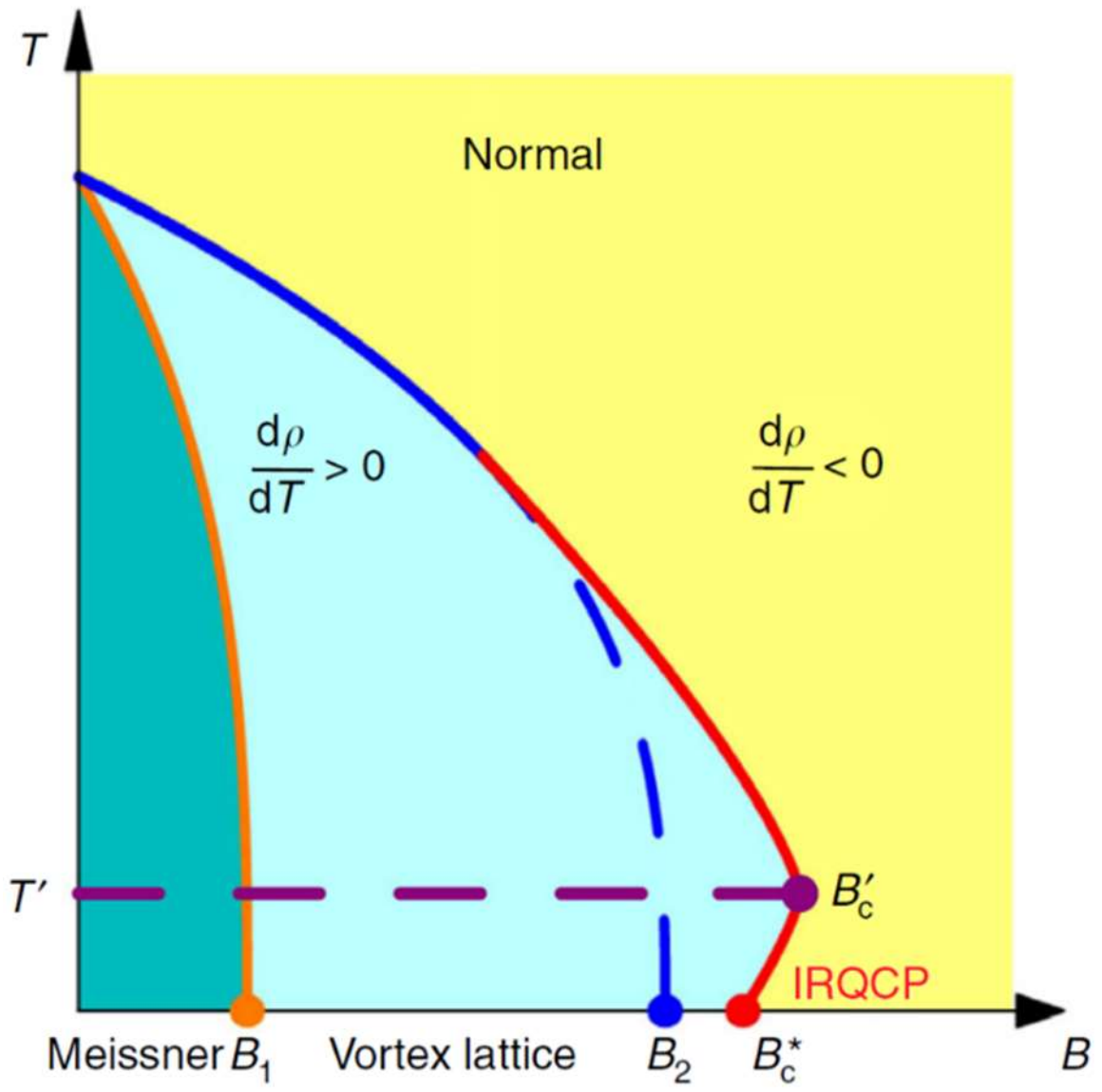

**Figure 9.** Schematic of the SMT phase diagram in the presence of strong fluctuation effect. Under the influence of superconducting fluctuations with strong SOI, the mean-field phase boundary (blue dashed line) buckles outward to the solid red line. The anomalous phase boundary exists in the regime between $B_c^*$ and $B_c'$. When approaching the infinite-randomness quantum critical point (IRQCP) $B_c^*$ along this anomalous phase boundary, the system exhibits anomalous QGS behavior. Reprinted from [107].

### 2.3.4 Quantum Griffiths singularity under parallel magnetic field

Under perpendicular magnetic field, the emergence of quantum Griffiths phase is often accompanied with a vortex lattice to vortex glass-like phase transition [14]. Under parallel field, the experimental investigation of QGS in the absence of magnetic field induced vortices is highly desired to clarify whether the vortex formation is critical to the emergence of QGS. In epitaxial 4-ML $PdTe_2$ films, Liu *et al.* reported the observation of QGS under both perpendicular and parallel magnetic fields [109]. The film under parallel field exhibits the characteristics of QGS similar to the observations under perpendicular field. Specifically, the magnetoresistance curves at various temperatures show a large transition region (Fig. 10(a)), and the critical fields of crossing points are in good agreement with the activated scaling theory (Fig. 10(b)). The critical exponent $z\nu$ obtained from finite size scaling analysis diverges as $z\nu \propto |B - B_c^*|^{-0.6}$ when approaching the quantum critical point (Fig. 10(c)). Furthermore, the activated scaling analysis with the irrelevant correction is performed to provide direct evidence of QGS (Fig. 10(d)). The details of the direct activated scaling will be introduced in Chapter 2.3.5. The detection of out-of-plane and in-plane QGS in the 4-ML $PdTe_2$ film indicates the universality of QGS under different field orientations. With increasing film thickness, the QGS disappears in the 6-ML film under perpendicular field but still appears under parallel field, suggesting different microscopic mechanism to form rare regions under different field directions. It is proposed that the disorder significantly influences the strength of SOI and thus the in-plane critical field, which gives rise to the quantum Griffiths phase in the absence of vortices [109]. The QGS in parallel magnetic field driven SMT has also been found in $\beta$-W films [116]. These observations may inspire further studies on quantum phase transitions under parallel magnetic fields and call for the microscopic understanding of quantum phase transitions in the absence of magnetic field induced vortices.

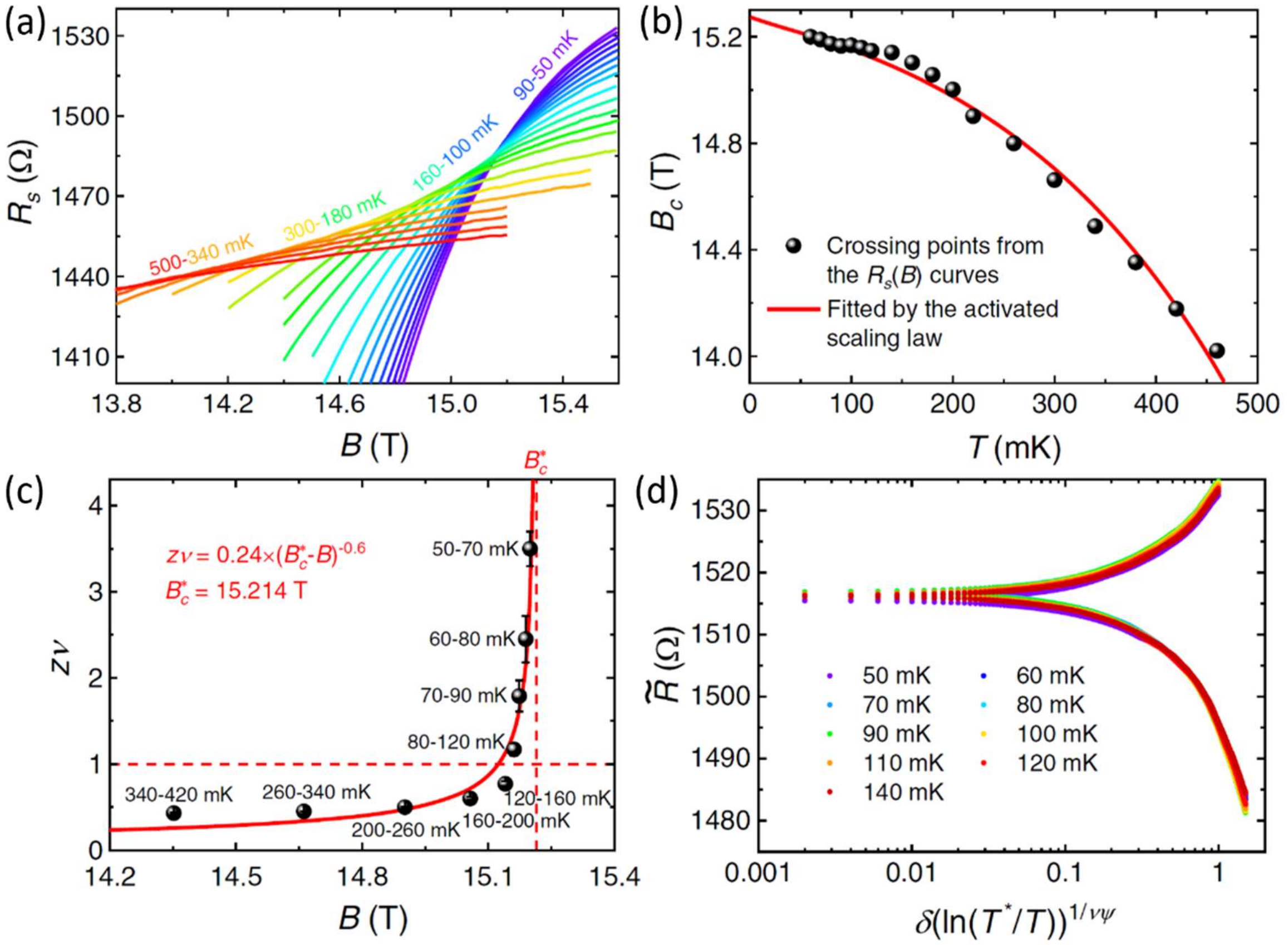

**Figure 10.** QGS under parallel magnetic field in 4-ML $PdTe_2$ film. (a) Parallel magnetic field dependence of sheet resistance at various temperatures. (b) Crossing points from the magnetoresistance isotherms. The solid red line is the fitting curve from the activated scaling analysis with the irrelevant correction. (c) The critical exponent $z\nu$ obtained from finite size scaling analysis diverges as $z\nu \propto |B - B_c^*|^{-0.6}$ when approaching the quantum critical point. (d) The direct activated scaling analysis of $\tilde{R}_s(B)$ curves ranging from 50 to 140 mK. $\tilde{R}_s$ denotes the sheet resistance considering the irrelevant correction. Reprinted from [109].

### 2.3.5 The direct activated scaling theory

So far, we have demonstrated that the finite size scaling analysis ($R = R_c F(\delta/T^{1/z\nu})$, where $\delta = |p - p_c|/p_c$ is the dimensionless parameter with the critical parameter $p_c$) provides a powerful tool to extract the critical exponent $z\nu$ in quantum phase transition [2-4, 12-14]. The activated scaling law gives rise to a divergent $z\nu$ ($z\nu \propto |\delta|^{-\nu\psi}$) when approaching the infinite-

randomness quantum critical point. The divergent $z\nu$ has been observed in SMT of diverse 2D superconducting systems [14, 107-112, 114-118] and identified as a primary feature of QGS.

Alternatively, it was proposed that the scaling dependence of resistance follows an activated scaling law with the irrelevant correction when approaching zero Kelvin [109, 113, 124]

$$R\left(\delta, ln\frac{T^*}{T}\right) = \Phi\left[\delta(ln\frac{T^*}{T})^{\frac{1}{\nu\psi}}, u(ln\frac{T^*}{T})^{-\frac{\omega}{\psi}}\right]. \qquad (6)$$

Here, $T^*$ is the characteristic temperature of quantum fluctuation, $u$ is the leading irrelevant scaling variable, and $\omega > 0$ is the irrelevant scaling exponent. In SMT of $PdTe_2$, $InO_x$ and $\beta$-W films, the direct activated scaling analysis is used to provide new evidence of QGS [109, 113, 116]. Figure 10(d) shows an example of the activated scaling analysis in 4-ML $PdTe_2$ film. Additionally, the critical field $B_c(T)$ of the phase boundary is given as $\frac{[B_c^* - B_c(T)]}{B_c^*} \sim u(ln\frac{T^*}{T})^{-\frac{1}{\nu\psi}-\frac{\omega}{\psi}}$, where $B_c^*$ is the characteristic field of the infinite-randomness quantum critical point. The SMT phase diagram of 4-ML $PdTe_2$ film, determined from the crossing points of adjacent $R_S(B)$ curves, has been shown in good agreement with the prediction of the activated scaling analysis (Fig. 10(b)) [109].

### 2.4 Discussion

As summarized in Table 3, the divergence of critical exponent $z\nu$ approaching the quantum critical point has been revealed in a wide variety of 2D superconducting systems. The experimental value of $z\nu$ in these systems increases significantly with decreasing temperature and does not have a trend to saturate even at the lowest temperature achievable in dilution refrigerator. Such behavior can be well explained in the framework of QGS.

The physical understanding of QGS is based on the existence of large rare regions in which the phase of order parameter is coherent. The time scale of phase evolution in rare region (*i.e.*, the temporal correlation length $\xi_\tau$) diverges rapidly when the size of rare region increases and approaches infinite value as $T \to 0$ K [97]. The slow dynamics of rare regions may dominate the critical behavior of global systems in certain circumstances and give rise to divergent critical exponent [97], which is a key feature of QGS. It is argued that the contribution of each rare region would need to diverge exponentially with its volume (or area in 2D) to lead to a true quantum Griffiths phase; conversely, the coupling to the heat bath can at most grow in proportion to the surface or edge of the rare region with a limited superconducting coherence length of $\xi_0$ [125]. However, the superconducting coherence length in BCS theory $\xi_{\mathrm{BCS}} = \frac{\hbar v_{\mathrm{F}}}{\pi\Delta(0)}$ [126] (where $\Delta(0)$ is the superconducting gap and $v_{\mathrm{F}}$ is Fermi velocity) diverges when the superconducting gap $\Delta(0)$ tends to zero approaching the quantum critical point. Moreover, an in-depth investigation of superconducting fluctuation in 2D system indicates the fluctuating Cooper pairs could form quantum liquid with long coherence length $\xi_{\mathrm{QF}} \gg \xi_{\mathrm{BCS}}$ close to the quantum critical point [127]. Therefore, the coupling to the heat bath would not be limited to the surface or edge of the rare region when approaching the quantum critical point, and hence the prerequisite of QGS is satisfied.

At the end of this chapter, we briefly discuss the relation between the critical behavior and dimensionality. We have presented the experimental observations of QGS in various 2D superconducting systems. The divergent critical exponent follows $z\nu \propto |\delta|^{-\nu}$ with $\nu\psi = 0.6$ approaching the infinite-randomness quantum critical point (where $\delta$ is the dimensionless parameter, $z$ is the dynamic critical exponent, $\nu$ is the correlation length critical exponent, and $\psi$ is the tunneling exponent), which is in good agreement with the numerical renormalization group results based on the random transverse-field Ising model [84, 85]. In 1D systems, the critical behavior $z\nu \propto |\delta|^{-\nu\psi}$ is also valid but with different critical exponent $\nu\psi = 1$ [83]. The experimental signature of QGS has been reported in a layered quasi-1D superconductor $Ta_2PdS_5$ [128]. The critical exponent diverges near the quantum critical point $B_c^*$ following $z\nu \propto |B - B_c^*|^{-\nu\psi}$ with $\nu\psi = 1.2$, which slightly deviates from the theoretical value of 1. In 3D systems, the numerical studies indicate the quenched disorder may also lead to an infinite-randomness quantum critical point. In the ferromagnetic to paramagnetic quantum phase transition of 3D alloy $Ni_{1-x}V_x$, the vanadium doping ($x$) dependence of exponents $\alpha$ and $1-\gamma$ follows the trend given by power-law QGS $\alpha(x) \propto |x - x_c|^{\nu\psi}$ and $1-\gamma(x) \propto |x - x_c|^{\nu\psi}$ with $\nu\psi = 0.34$ near the quantum critical point $x_c$ (Fig. 5) [90, 100]. So far, the QGS in 3D superconductors has not been reported. Related experimental and theoretical studies are necessary to demonstrate the universality of QGS in 3D SIT/SMT and ferromagnet-paramagnet transitions, as well as the relation between the critical behavior and dimensionality.

## 3. Anomalous metal state

SIT/SMT as a paradigm of quantum phase transition has been widely investigated over the last three decades. It was long believed that the bosons would either localize as the insulating state or condense as the superconducting state in 2D systems based on the Heisenberg uncertainty principle. However, this theoretical scenario has been challenged by the experimental observation of possible intervening metallic state in the SIT/SMT, where the resistance drops with decreasing temperature below $T_c$ and saturates to a value far below the normal state resistance [125]. The intervening metallic state might represent an unexpected quantum ground state approaching 0 K for 2D bosonic systems, which is called as the quantum metal state [33], anomalous metal state [125] or the Bose metal [129]. The bosonic nature of the metallic ground state will be illustrated in Chapter 3.2.2.

### 3.1 The intervening metallic states in the quantum phase transitions

#### 3.1.1 Experimental signatures in strongly disordered 2D superconductors

Signatures of anomalous metal states have been widely reported in the investigations of disordered superconducting systems [125, 130-147], which could be realized by tuning control parameters such as film thickness, magnetic field and gate voltage. An early signature of anomalous metal state was revealed by Jaeger *et al.* [131]. The evolution of temperature dependent sheet resistance $R(T)$ with different film thickness exhibits disorder tuned SIT in ultrathin granular metal films of Al, In, Ga and Pb (Fig. 11). With decreasing temperature, the sheet resistance saturates to a value far below the normal state resistance at low temperatures, reminiscent of the metallic characteristic. This metallic state disrupts the direct transition from the superconducting state to the insulating state, indicating an intervening ground state.

Experimental signature of anomalous metal state was also observed in the magnetic field induced SMT in α-MoGe thin films [132-134]. Figure 12 shows the sheet resistance plotted in the logarithmic scale as a function of $1/T$. In the high temperature regime, $\lg(R)$ follows a linear trend with respect to $1/T$, indicating the behavior of thermally activated flux flow. The resistance decreases with decreasing temperature and saturates to a finite value at lower temperatures, suggesting the appearance of a metallic state. The saturated resistance at low temperatures in the presence of magnetic field could be interpreted as a result of the coupling to dissipative environment [148].

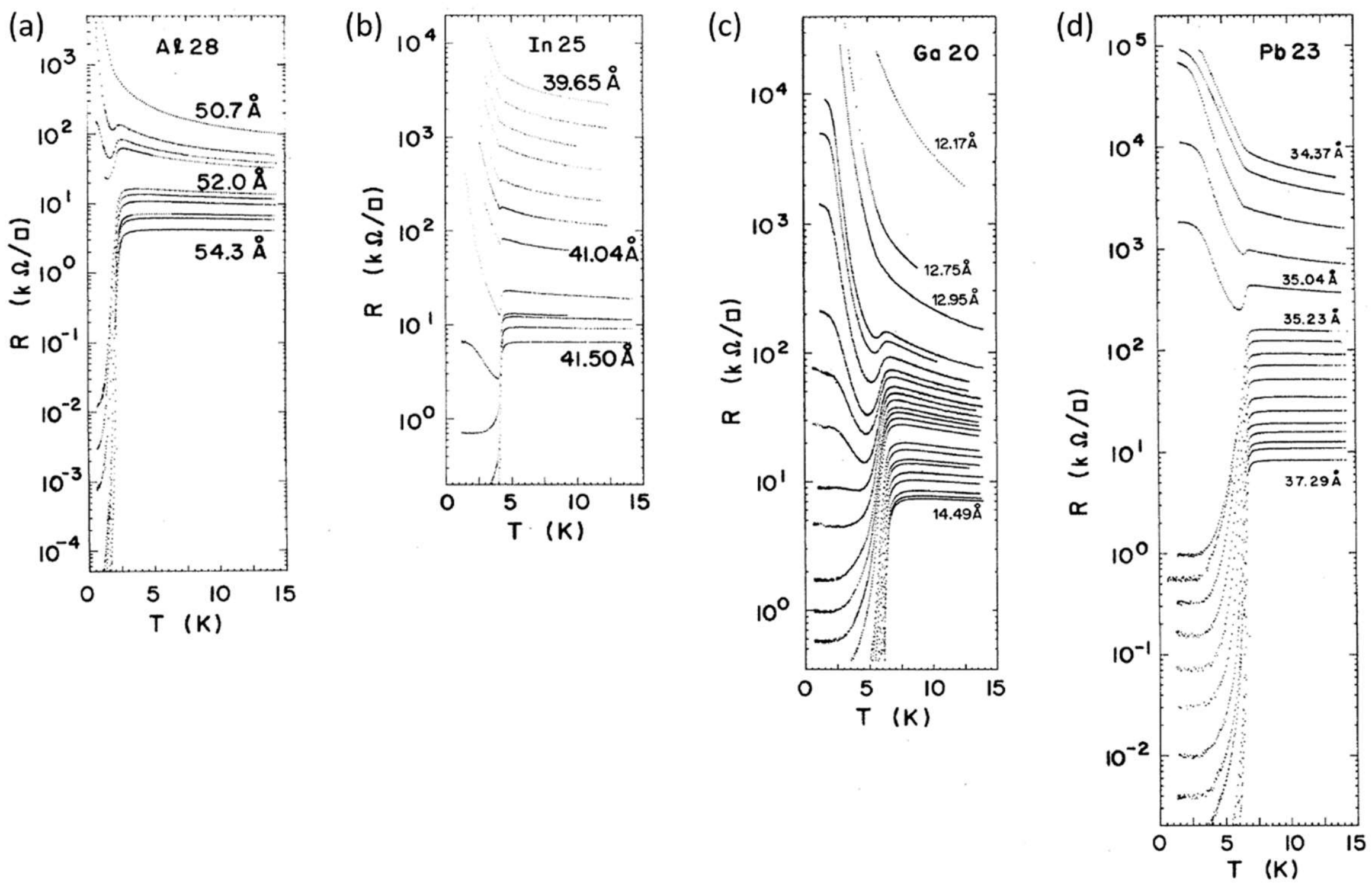

**Figure 11.** The evolution of $R(T)$ curves with respect to film thickness in (a) Al, (b) In, (c) Ga, and (d) Pb films. The resistance saturates at low temperature to a value that is far below the normal state resistance. Reprinted from [131].

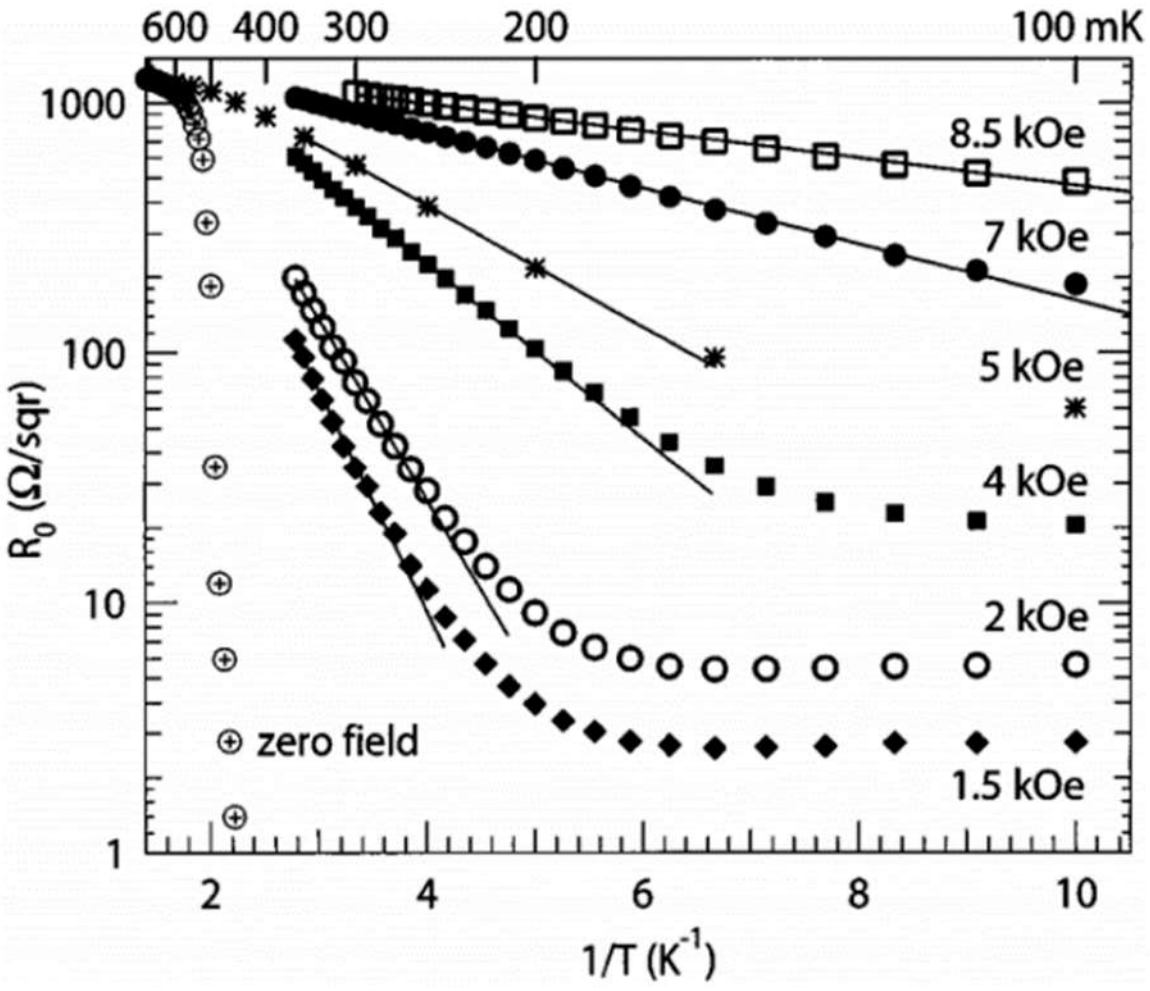

**Figure 12.** $R(T)$ curves plotted in the logarithmic scale as a function of $1/T$ at various magnetic fields in an α-MoGe film. The straight lines show the activated behavior in the high temperature region; the resistance saturates to a value at lower temperatures. Reprinted from [132].

The resistance saturation in the low temperature regime has also been reported in magnetic field induced SMTs in other disordered superconducting thin films, such as Ta films [135], $TaN_x$ and $InO_x$ films [136]. Besides, similar phenomena have been observed in gate voltage driven SMTs in artificially fabricated superconducting islands arrays, including Sn disks array on graphene substrate [138] and Al islands array coupled with InAs quantum well [139]. In addition to the resistance saturation at low temperatures, the intervening metallic state shows other experimental features such as vanishing Hall resistance [136] and absence of cyclotron resonance [140, 141], indicating an emergent particle-hole symmetry.

### 3.1.2 Recent progress in highly crystalline 2D superconductors

The investigations on disordered superconducting systems have revealed signatures of anomalous metal state, characterized by the saturated sheet resistance at low temperatures. The developments of film growth and nanofabrication technique in recent decades have made new opportunity to study the quantum phase transitions and the anomalous metal states in crystalline superconducting films and nanodevices. In this section, we focus on recent progress on the signatures of anomalous metal states observed in 2D crystalline superconductors [32, 38, 149-162].

Saito *et al.* investigated the magneto-transport properties of the ionic-liquid gated superconductivity in crystalline ZrNCl nanodevices [38]. At gate voltage of $V_g = 6.5$ V, a metallic state with saturated resistance was observed at low temperatures by applying perpendicular magnetic field (Fig. 13(a)). The sheet resistance $R_{\mathrm{sheet}}$ exhibits an activated behavior below $T_c$ described by $R_{\mathrm{sheet}} = R'\exp{(-U/k_BT)}$, where $U$ is the activation energy and $k_B$ is the Boltzmann constant. The extracted activation energy $U/k_B$ follows the trend given by $U = U_0 \ln{(H_0/H)}$ (Fig. 13(b)), suggesting the thermal collective vortex creep in the 2D system [163]. At lower temperatures, $R_{\mathrm{sheet}}$ deviates from the activated behavior and then saturates to a value down to the lowest measurement temperature even at a small magnetic field of 0.05 T. The field dependence of the saturating resistance is consistent with the quantum creep model in the strong dissipation limit [148]

$$R_{\mathrm{sheet}} \sim \frac{\hbar}{4e^2}\frac{\kappa}{1-\kappa}, \text{ with } \kappa \sim \exp{[C\frac{\hbar}{e^2}\frac{1}{R_{\mathrm{n}}}\left(\frac{H-H_{c2}}{H}\right)]}. \quad (7)$$

Here, $C$ is a dimensionless constant, $R_{\mathrm{n}}$ is the sheet resistance of the normal state, and $H_{c2}$ is the perpendicular upper critical field. Besides ZrNCl, experimental signatures of the anomalous metal states have also been reported in other crystalline nanodevices made of exfoliated 2D materials, such as $WTe_2$ [149], $TiSe_2$ [150], $MoS_2$ [151], $TaSe_2$ [152], $Li_xTiSe_2$ [153], and $WS_2$ [154] (see Table 4 for a list of representative anomalous metal states and the tuning methods).

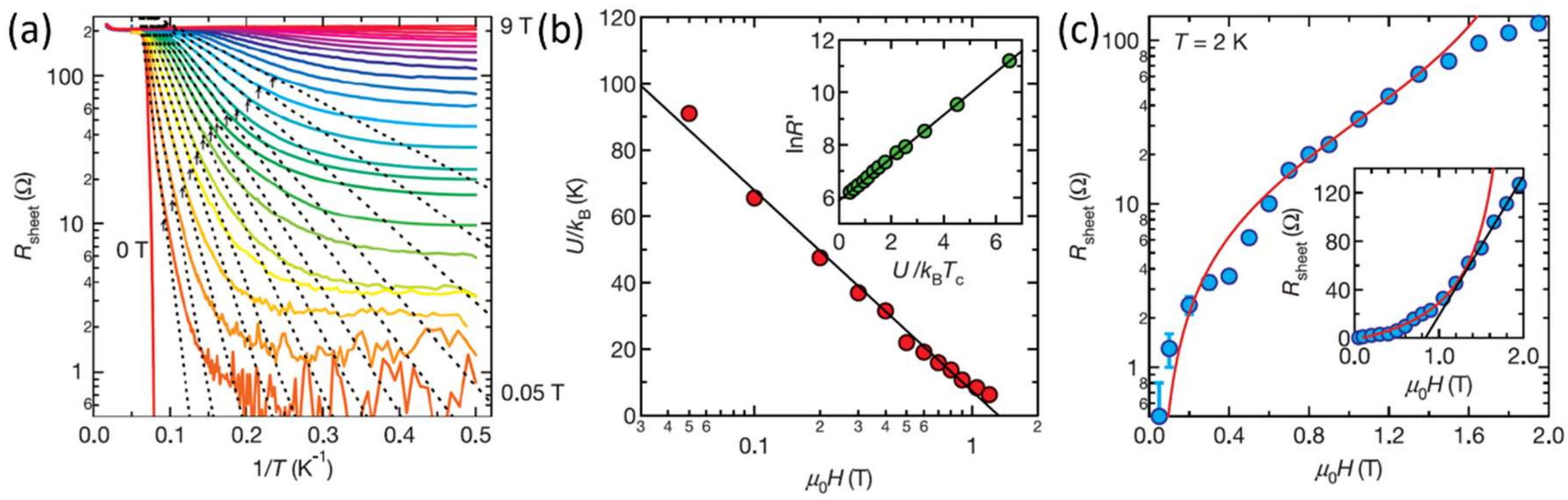

**Figure 13.** (a) Sheet resistance $R_{\mathrm{sheet}}$ plotted in the logarithmic scale as a function of $1/T$ at various magnetic fields in ion-gated ZrNCl nanoflake at $V_g = 6.5$ V. The black dashed lines

describe the thermally activated behavior by $R_{\text{sheet}} = R'\exp(-U/k_B T)$. (b) The activation energy $U/k_B$ as a function of magnetic field follows the trend given by $U = U_0 \ln(H_0/H)$ (black solid line), suggesting the 2D thermal collective vortex creep in the ion-gated 2D superconducting ZrNCl. Inset, the relation between $\ln(R')$ and $U/k_B T_c$. (c) The saturating resistance at low temperature as a function of magnetic field at 2 K. The red solid line is a fit using the quantum creep model. Inset, $R_{\text{sheet}}$ shows a linear field dependence above 1.3 T (black solid line). Reprinted from [38].

Crystalline thin films and interfaces provide another platform to investigate the anomalous metal state. At the $LaAlO_3/SrTiO_3$ (001) interface, resistance saturation as a signature of the anomalous metal state was observed below 100 mK in the top-gate voltage tuned SIT (Fig. 14) [158]. Dual electrostatic gates were effective to tune the resistance at 40 mK ranging from far below to far above the quantum resistance $h/4e^2$. The saturated resistance at low temperatures has also been revealed in gate voltage tuned SIT in other oxide thin films and interfaces, such as $La_2CuO_{4+\delta}$ thin films [157] and $LaAlO_3/KTaO_3$ (111) interfaces [160]. Moreover, the signatures of the anomalous metal state have been reported in atomically flat crystalline films grown by MBE, such as $PdTe_2$ thin films [161], ultrathin crystalline FeSe films grown on $SrTiO_3$ (FeSe/STO) [32], and mono-unit-layer $NbSe_2$ films [162].

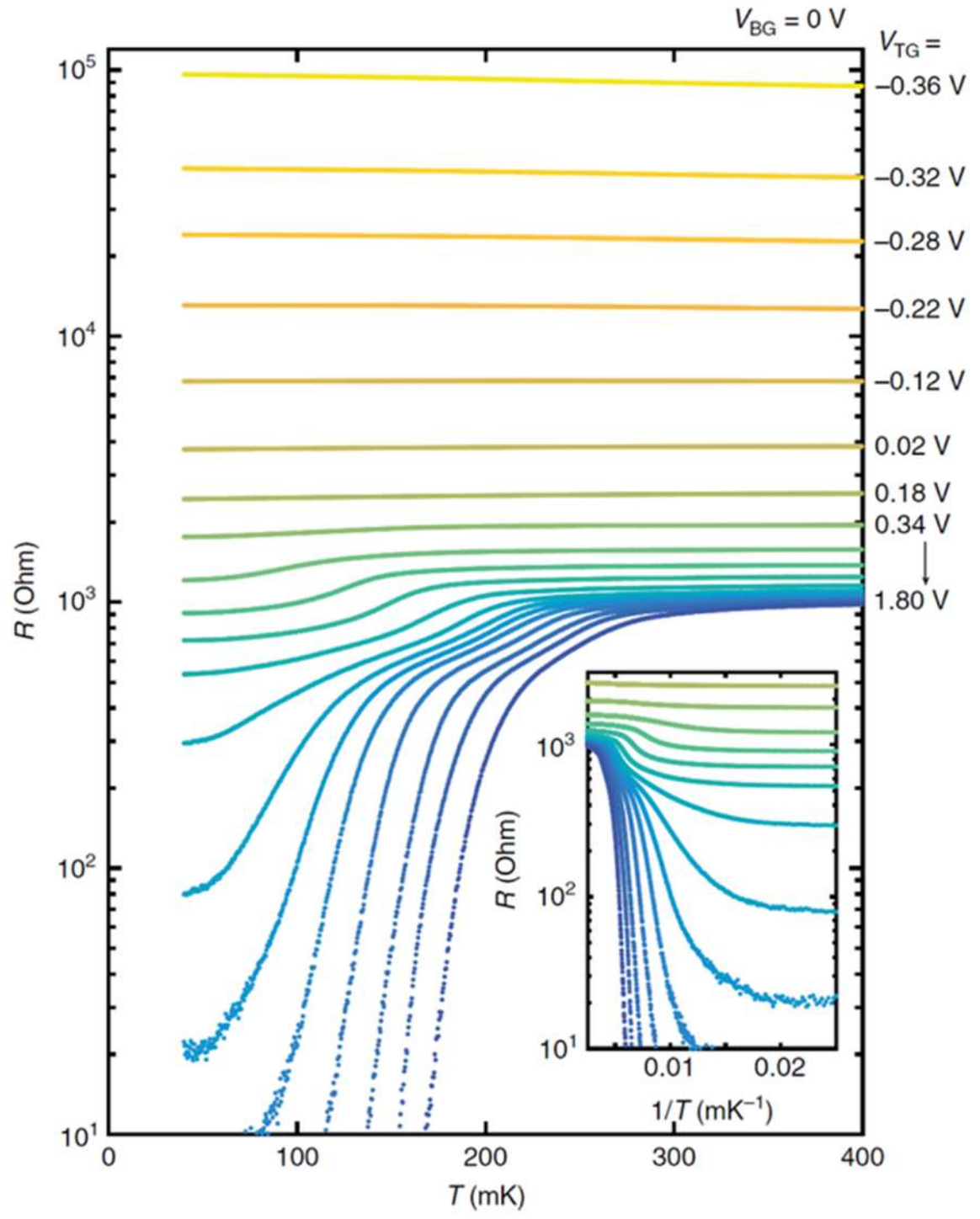

**Figure 14.** Temperature dependent sheet resistance at $LaAlO_3/SrTiO_3$ (001) interface at various top-gate voltages ranging from -0.36 to 1.80 V with back-gate voltages fixed at 0 V. Inset, sheet resistance plotted with respect to $1/T$ shows resistance saturation behavior at low temperatures. Reprinted from [158].

As a summary of the section, the signature of anomalous metal state has been reported in various 2D superconducting systems, from strongly disordered superconducting systems to highly crystalline superconducting systems. A list of representative anomalous metal states reported in 2D superconductors is summarized in Table 4. While the disorder may be critical in the quantum phase transition, the appearance of anomalous metal state does not seem to depend on the detailed morphology of the disorder [125].

**Table 4.** Experimental signatures of anomalous metal states reported in various 2D superconducting systems.

| System | Description of the system | Characteristics of the anomalous metal state | Reference |
|---|---|---|---|
| **Al, In, Ga and Pb thin films** | Granular metal thin films | Resistance saturation at low temperatures in thickness tuned SIT. | 131 |
| **α-MoGe thin films** | Amorphous thin films | Resistance saturation is observed at low temperatures by applying perpendicular magnetic field. | 132 |
| **$InO_x$ thin films** | Amorphous oxide thin films | Resistance saturation is observed at low temperatures by applying perpendicular magnetic field. The metallic state is characterized by vanishing Hall resistance and absence of cyclotron resonance. | 136, 141 |
| **Sn disks array on graphene, Al islands array coupled with InAs quantum well** | Artificial arrays of superconducting islands on 2D conductors | Resistance saturation is observed at low temperatures by applying gate voltage. | 138, 139 |
| **ZrNCl nanoflakes** | Exfoliated 2D superconductor tuned by ionic liquid gate | Resistance saturation is observed at low temperatures by applying perpendicular magnetic field. | 38 |
| **$WTe_2$ monolayer** | Induced superconductivity in a monolayer topological insulator tuned by gate voltage | Resistance saturation is observed at low temperatures by applying gate voltage and magnetic field. | 149 |
| **1T-$TiSe_2$ nanoflakes** | Exfoliated 2D superconductor tuned by ionic liquid gate | Resistance saturation is observed at low temperatures by applying perpendicular magnetic field. | 150 |
| **$MoS_2$ nanoflakes** | Exfoliated 2D superconductor tuned by ionic liquid gate | Resistance saturation is observed at low temperatures by applying perpendicular magnetic field. | 151 |
| **4Ha-$TaSe_2$ nanodevices** | Exfoliated 2D superconductor from single crystals | In the 7 nm- thick nanodevice, resistance saturation is observed at low temperatures by applying perpendicular magnetic field. | 152 |
| **Lithium intercalated $TiSe_2$ nanoflakes** | Exfoliated 2D superconductor from single crystals | Resistance saturation is observed at low temperatures by applying perpendicular magnetic field. | 153 |
| **Few-layer 1T'-$WS_2$** | Exfoliated 2D superconductor from single crystals | Resistance saturation is observed at low temperatures by applying perpendicular magnetic field. | 154 |
| **Nano-patterned YBCO thin films** | Nanopore-modulated oxide thin films (reactive ion etching using AAO membrane mask) | Resistance saturation is observed at low temperatures in SIT tuned by increasing etching time. The onset temperature of the anomalous metal state $T_{\mathrm{AM}}$ is up to 10 K. The bosonic nature of the anomalous metal state is evidenced by charge-2$e$ quantum oscillations and vanishing Hall resistance. | 156 |
| **$La_2CuO_{4+\delta}$ thin films** | Oxide thin films grown by oxide MBE | Resistance saturation is observed at low temperatures by applying gate voltage. The onset temperature of the anomalous metal state $T_{\mathrm{AM}}$ is up to 10 K. | 157 |
| **$LaAlO_3/SrTiO_3$ interfaces** | Oxide interfaces grown by PLD | Resistance saturation is observed at low temperatures by applying gate voltage. | 158 |
| **$LaAlO_3/KTaO_3$ interfaces** | Oxide interfaces grown by PLD | Resistance saturation is observed at low temperatures. | 160 |
| **$PdTe_2$ thin films** | Atomically flat crystalline film grown by MBE | Resistance saturation is observed at low temperatures by applying perpendicular magnetic field. | 161 |
| **$FeSe/SrTiO_3$** | Atomically flat crystalline film grown by MBE | Resistance saturation is observed at low temperatures in both pristine and nanopatterned films. The onset temperature of the anomalous metal state $T_{\mathrm{AM}}$ is up to 20 K. | 32 |
| **Mono-unit-layer $NbSe_2$** | Atomically flat crystalline film grown by MBE | In situ transport measurements show resistance saturation at low temperatures in perpendicular magnetic field. | 162 |

### 3.1.3 Extrinsic origin of metallic states in fragile 2D superconductors

Resistance saturation as a signature of the anomalous metal state has been reported in a variety of 2D superconducting systems. However, there are intensive debates on whether the resistance saturation at low temperature is due to extrinsic effects such as electromagnetic noise and carrier overheating. Thus, it is crucial to distinguish the intrinsic metallic states from those with the extrinsic origin.

It is argued that the external electromagnetic noise in the measurements may disrupt the phase coherence in 2D superconducting systems, which casts doubt on the intrinsic origin of the anomalous metal state. Tamir *et al.* investigated the metallic states at low temperatures in two different thin-film superconductors, amorphous $InO_x$ thin film and exfoliated 2H-$NbSe_2$ nanosheet [164]. Figure 15 displays the resistance curves as a function of $1/T$ obtained with and without filters. In both samples, the red curves measured without filters show a signature of the anomalous metal state, where the resistance initially decreases with decreasing temperature following an activated behavior and then saturates to a value at lower temperatures. However, after installing the external low-pass filters, the resistance follows the activated behavior down to ultralow temperatures and the resistance saturation behavior disappears (the blue curves), indicating that the metallic behavior could be eliminated by adequately filtering the external noise [164]. The so-called "anomalous metal state" induced by the external electromagnetic radiation was also reported in superconducting Ta thin films [165].

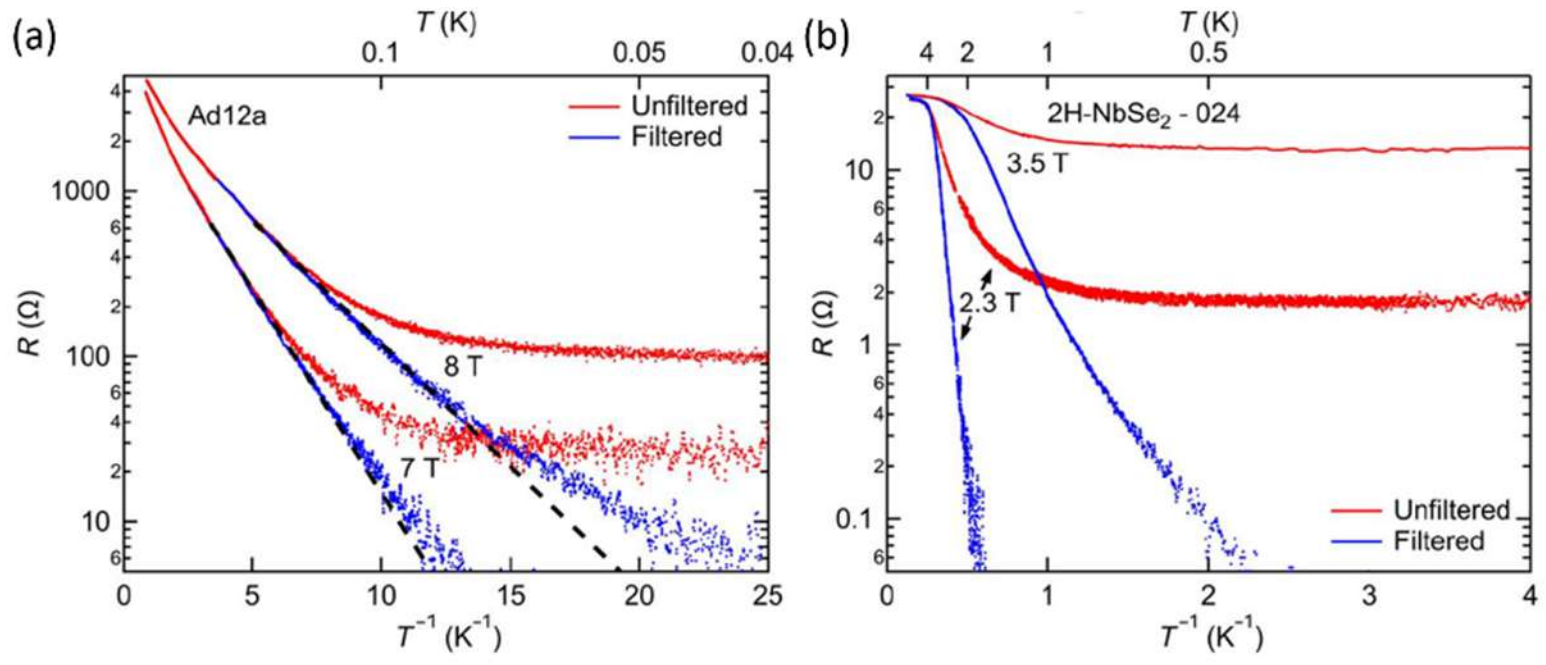

**Figure 15.** (a) $R(T^{-1})$ curves obtained from amorphous $InO_x$ thin films at 7 and 8 T. The black dashed lines indicate an activated behavior. (b) $R(T^{-1})$ curves obtained from ultrathin crystalline $NbSe_2$ devices at 2.3 and 3.5 T. In (a) and (b), the blue curves are the measurement results with filters while the red curves are the measurement results without filters. Reprinted from [164].

Carrier overheating is another extrinsic factor that may give rise to the resistance saturation at ultralow temperature. Either the measuring current or the current result from the electromagnetic noise could bring the dissipation into the system. Since the electronic heat capacities of ultrathin films are very small, the dissipation may prevent cooling down the electrons in the film at millikelvin temperatures [166]. The role of heating from the measuring current has been shown in amorphous Bi films [166] and exfoliated $NbSe_2$ nanodevices [167]. Parendo *et al*. found that the resistance saturation behavior below 200 mK in amorphous Bi films was due to the failure to cool down the electrons in the film [166]. This indicates that the resistance saturation at ultralow temperatures may not be intrinsically metallic but have external origin. Benyamini *et al.* performed the non-local measurements (see Fig. 16(a) for a schematic) in exfoliated $NbSe_2$ nanodevices with different d.c. source currents (Fig. 16) [167]. With low d.c. source current, the differential resistances measured by all non-local probes show the activated behavior. With higher d.c. source current, the area closest to the source-drain electrodes shows a saturated behavior while the furthest area still shows the activated behavior. With the highest d.c. source current, the source-drain area is in the normal state while the other regions show the saturated behavior. The non-local measurements demonstrate the evolution from the dissipationless to dissipative transport with increasing measuring current.

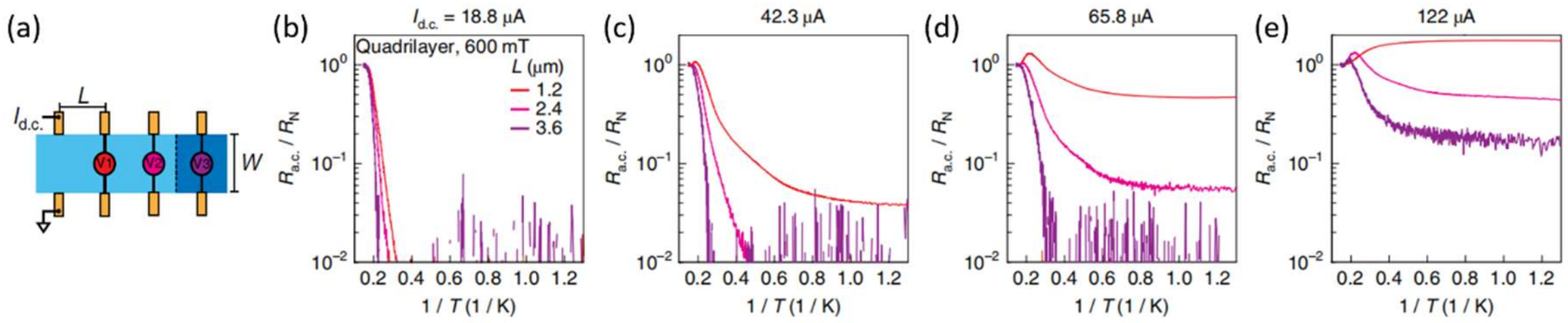

**Figure 16.** (a) A schematic for non-local measurements. The measurements were taken at different distances ($L$) from the source-drain electrodes. (b-e) Normalized non-local differential resistance in exfoliated $NbSe_2$ nanodevices. The measurements were taken at four d.c. currents of 18.8 μA, 42.3 μA, 65.8 μA, and 122 μA. Reprinted from [167].

## 3.2 Experimental evidence and main characteristics of anomalous metal state

### 3.2.1 Detection of anomalous metal state with high quality filters

As mentioned in Chapter 3.1.3, the resistance saturation at low temperatures could be due to the extrinsic effects such as electromagnetic noise and carrier overheating, which casts doubt on the intrinsic origin of the anomalous metal state. In order to obtain evidence of the intrinsic metallic state, it is crucial to perform measurements with high-quality filters to minimize the influence of electromagnetic noise. Besides, it is important to confirm that the resistance at low temperature is independent on the measuring current to exclude the effect of carrier overheating.

Yang *et al.* systematically investigated the anomalous metal states in high-temperature superconducting YBCO films [156]. The 12-nm-thick YBCO films were patterned with a triangular array of nanoholes using reactive ion etching (RIE) through anodized aluminum oxide (AAO) membrane mask. The films undergo the SIT with increasing etching time, and the anomalous metal state is detected between the superconducting and insulating regimes. The resistance saturation at low temperatures as the experimental signature of anomalous metal state was detected in a series of nanopatterned YBCO films (Fig. 17(a)). With or without using resistor-capacitor (RC) filters in the measurements, $R_s(T)$ curves show similar saturating behaviors down to 50 mK (Fig. 17(b)), excluding the possibility that the observed metallic states are originated from the external electromagnetic noise. Besides, the measurement current for the anomalous metal state of the YBCO film in Fig. 17(b) is no larger than 100 nA and $I-V$ characteristic of the anomalous metal state at 50 mK shows a good linearity up to 100 nA (Inset of Fig. 17(b)), indicating the resistance is independent on the excitation current. The extrinsic effects such as electromagnetic noise and carrier overheating are thus excluded, confirming the intrinsic origin of the anomalous metal state in nanopatterned YBCO films.

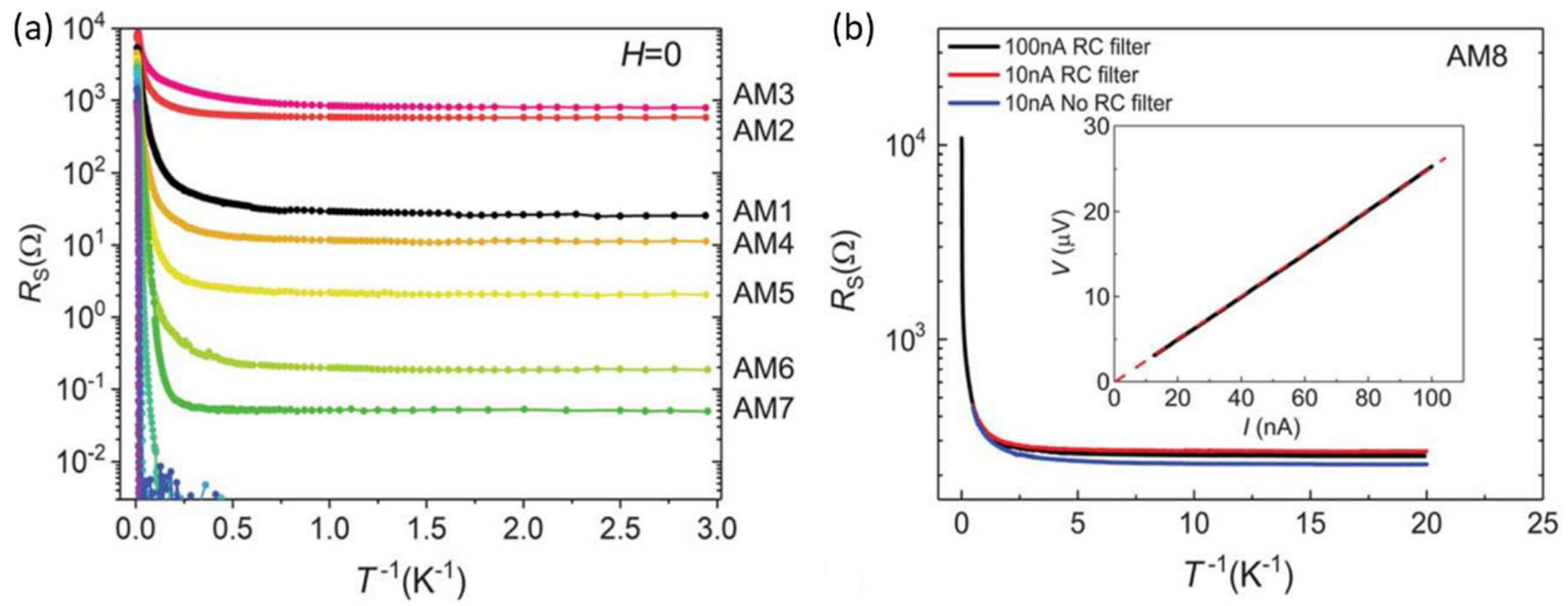

**Figure 17.** (a) Arrhenius plots of $R_s(T)$ curves showing the appearance of the anomalous metal states in nanopatterned YBCO films. (b) $R_s(T)$ with and without RC filters measured at low

temperatures down to 50 mK. Inset, $I-V$ characteristic of the anomalous metal state shows a good linearity at 50 mK. Reprinted from [156].

For ultralow temperature measurements in dilution refrigerators, effective filtering is indispensable to minimize high-frequency noise and thus avoid the electron temperature elevating significantly from the lattice temperature. The silver-epoxy filter, RC filter, and their combination are commonly used in the ultralow temperature transport measurements [152, 161, 168]. A typical measurement setup with silver-epoxy filters and RC filters is shown in Fig. 18(a), where the silver epoxy filters (installed at the low-temperature sample stage) and the RC filters (installed outside the cryostat) are connected in series with each lead of the sample [152]. The high-quality silver-epoxy filter, which takes the advantage of the skin effect, has an attenuation of more than 90 dB at high frequency above 600 MHz. The RC filter has an effective attenuation from 0.3 to 10 MHz but has a weaker attenuation at higher frequency. The combination of silver-epoxy filter and RC filter provides a better performance (Fig. 18(b)) [168].

Although the experimental signatures of anomalous metal states have been reported in various 2D superconductors, the evidences of anomalous metal states in MBE-grown high quality 2D superconducting films are scarce. By utilizing high-quality filters at ultralow temperatures, Liu *et al.* systematically studied the anomalous metal states in MBE-grown crystalline superconducting $PdTe_2$ thin films [161]. Under perpendicular magnetic field, the resistance drops below $T_c$ with an activated behavior in the higher temperature regime and then saturates to a finite value at lower temperatures. The measurements are performed using the well-filtered electrical leads. The combination of the thermocoax filter and the RC filter provides a good performance (Fig. 18(c) and (d)). The attenuation in the high frequency range (0.3 MHz to 10 GHz) is better than -50 dB, which indicates that more than 99.999% energy of the high-frequency noise has been reduced. The filter has an even better performance and reaches -100 dB above 1.1 GHz. The measurements with the high-quality filters can safely exclude the extrinsic origin of resistance saturation due to high-frequency noise and hence provide solid evidence of the anomalous metal state. Besides, the $I-V$ characteristic has a linear dependence, indicating the resistance is independent on the measuring current and the detected anomalous metal state in crystalline superconducting $PdTe_2$ thin films is intrinsic.

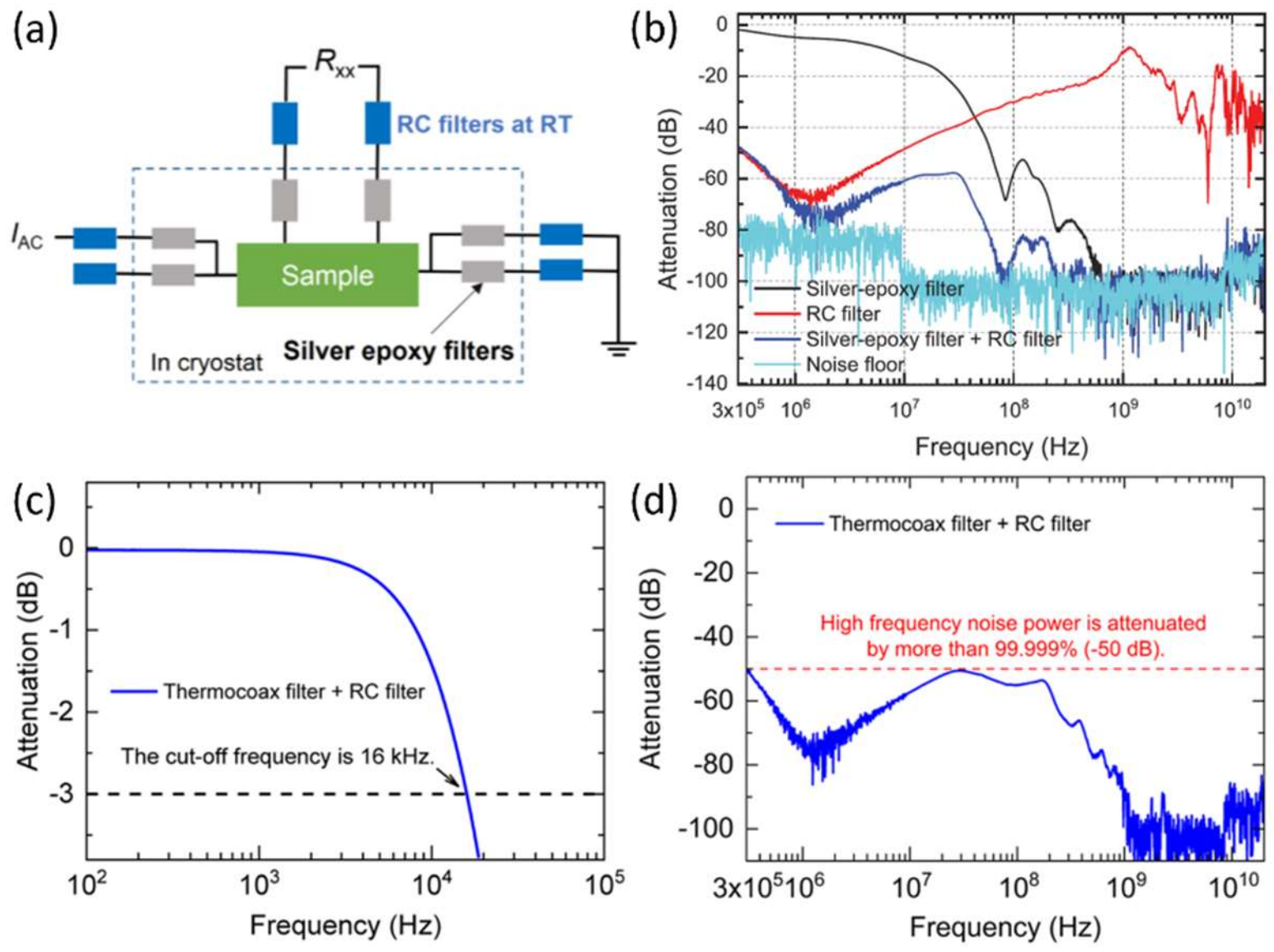

**Figure 18.** (a) Typical ultralow temperature measurement setup with silver-epoxy filters and RC filters. Reprinted from [152]. (b) Performance of silver-epoxy filter, RC filter and their combination. Reprinted from [168]. (c) and (d) Performance of the combination of thermocoax filter and RC filter. The cutoff frequency is around 16 kHz. The performance becomes better at higher frequency and reaches -100 dB above 1.1 GHz. Figure (c) and (d) are reprinted from [161].

Furthermore, Xing *et al*. revealed that the intrinsic anomalous metal states can still exist in 2D transition metal dichalcogenide superconducting devices despite the sensitivity of nanodevices to external noise. The exfoliated 4Ha-$TaSe_2$ superconducting devices were carefully studied with high-quality filters [152]. In a 5.0 nm-thick $TaSe_2$ nanodevice, the resistance saturation is observed at low temperatures down to 0.55 K and perpendicular fields of 0.1, 0.2 and 0.3 T in the unfiltered measurements. However, the saturation behavior disappears in the measurement setup with filters (Fig. 19(a)), indicating the observed “anomalous metal state” is extrinsic. However, at lower temperatures and higher magnetic fields, the intrinsic anomalous metal state emerges in a 7.0 nm-thick $TaSe_2$ nanodevice (Fig. 19(b)). The resistance saturation at ultralow temperatures down to 55 mK is detected in a dilution refrigerator equipped with the high-quality silver-epoxy and RC filters. The excitation current used in the measurement is 10 to 20 nA, which is within the linear $I-V$ regime of the observed metal states. The extrinsic factors such as high frequency noise and large excitation current are thus excluded, confirming the intrinsic origin of the anomalous metal state. The intrinsic anomalous metal state is further confirmed in another 4.4 nm-thick $TaSe_2$ nanodevice [152]. Figure 19(c) shows the phase diagram of 4Ha-$TaSe_2$ superconducting devices, displaying both extrinsic and intrinsic metallic states. Compared to the extrinsic anomalous metal state, the intrinsic anomalous metal state detected with high-quality filters appears in lower temperature region approaching zero temperature and higher magnetic field, indicating a quantum ground state. Table 5 summarizes experimental results reporting anomalous metal states in 2D superconductors which may exclude the effect of electromagnetic noise and (or) carrier overheating.

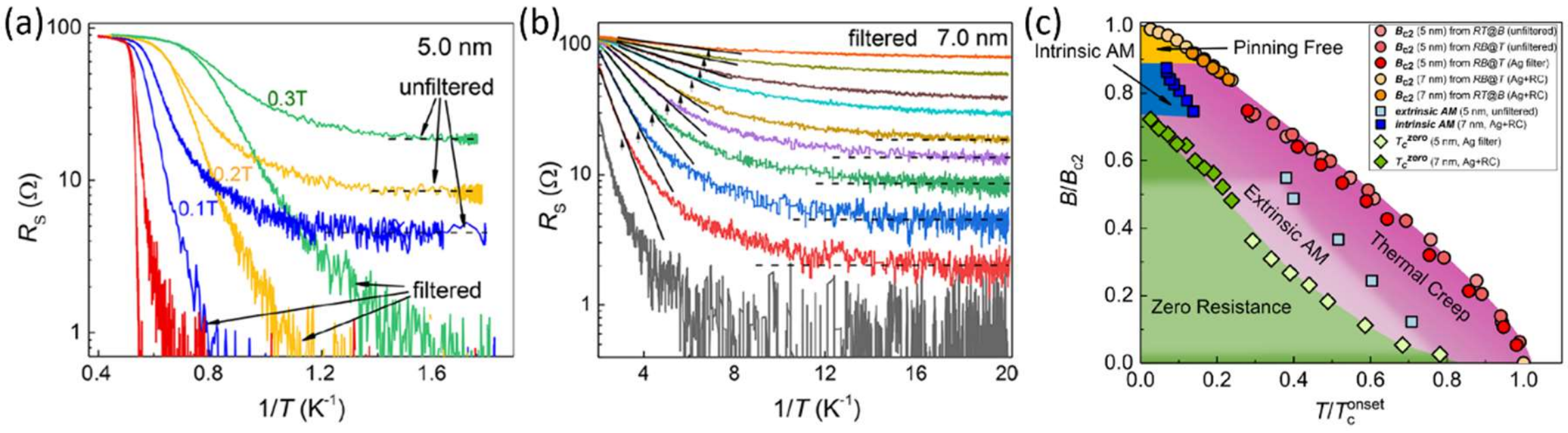

**Figure 19.** (a) The filtered and unfiltered resistance measurement results in the perpendicular magnetic field, suggesting extrinsic anomalous metal states in a 5.0 nm-thick 4Ha-$TaSe_2$ nanodevice at low temperatures down to 0.55 K. (b) Intrinsic anomalous metal states in a 7.0 nm-thick 4Ha-$TaSe_2$ nanodevice revealed by filtered measurements at ultralow temperatures down to 55 mK and perpendicular magnetic fields ranging from 0.704 to 1.254 T. (c) Phase diagram of 2D $TaSe_2$. The blue region and white transparent region represent the intrinsic and extrinsic anomalous metal states, respectively. Reprinted from [152].

**Table 5.** Representative works reporting anomalous metal states in 2D superconducting systems which may exclude the effect of electromagnetic noise and (or) carrier overheating.

| System | Exclusion of electromagnetic noise | Exclusion of carrier overheating | Reference |
|---|---|---|---|
| **Nano-patterned YBCO thin films** | With or without using RC filters in the measurements, $R(T)$ curves show similar saturating behaviors down to 50 mK. | $I-V$ curve has a linear dependence within 100 nA down to 50 mK; the measurement current for anomalous metal states down to 50 mK is within the linear $I-V$ regime. | 156 |
| **$PdTe_2$ thin films** | The ultralow temperature measurements are performed using a combination of the thermocoax filter and the RC filter. | $I-V$ curve has a linear dependence within 8 nA down to 15 mK; the measurement current for anomalous metal states down to 15 mK is within the linear $I-V$ regime. | 161 |
| **4Ha-$TaSe_2$ nanodevices** | The anomalous metal state is detected in a dilution refrigerator equipped with high-quality silver-epoxy filters and RC filters. | The excitation current used in the measurement is 10 to 20 nA, which is within the linear $I-V$ regime of the anomalous metal states. | 152 |
| **FeSe/$SrTiO_3$** | Radio frequency filters are used in the ultralow temperature measurements. | $I-V$ curve has a linear dependence within 40 nA down to 50 mK; the measurement current for anomalous metal states down to 50 mK is within the linear $I-V$ regime. | 32 |
| **1T-$TiSe_2$ nanoflakes** | Low-pass frequency filters are used to suppress high-frequency noise. | | 150 |
| **Lithium intercalated $TiSe_2$ nanoflakes** | $\pi$ filters are used to reduce high-frequency noise. | | 153 |
| **$MoS_2$ nanoflakes** | The heating effect from extrinsic noise can be excluded by detecting suppressed second harmonic signal in the anomalous metal state. | | 151 |
| **Granular In/$InO_x$ composite** | The measurements are filtered using $\pi$ filters and cryogenic filters. | Linear response to excitation current is confirmed at various temperatures and magnetic fields. | 146, 147 |

### 3.2.2 The bosonic nature of anomalous metal state

It was long believed that the bosons would only stay in either the superconducting or insulating state. According to the Heisenberg uncertainty principle, the particle number and phase as conjugate variables cannot be determined simultaneously. As a result, the bosons can either be in an eigenstate of particle number (*i.e.*, the insulating state) or an eigenstate of phase (*i.e.*, the superconducting state) [129]. Thus, the detection of anomalous metal states in SIT/SMT challenged the traditional theory and attracted tremendous interests in both experimental and theoretical studies. Although multiple physical scenarios have been developed to describe the anomalous metal behavior, the nature of anomalous metal state in 2D systems still remains unclear.

Yang *et al.* systematically studied the evolution of quantum phase coherence across the superconductor-metal-insulator transition through magnetoconductance quantum oscillations in nanopatterned high-temperature superconducting YBCO films [156]. The films undergo the SIT with increasing etching time, and the intervening anomalous metal state is detected across the transition at low temperatures. The longitudinal resistance ($R_{xx}$) and the Hall resistance ($R_{xy}$) are measured simultaneously to investigate the nature of the anomalous metal state. At low temperatures, $R_{xx}(T)$ saturation is detected and the slope of $R_{xy}(B)$ is observed to drop to zero (Fig. 20(a)). The vanishing Hall resistance was ever noticed by Breznay and Kapitulnik in $InO_x$ and $TaN_x$ films as experimental indication of the emergent particle-hole symmetry in the anomalous metal state [136].

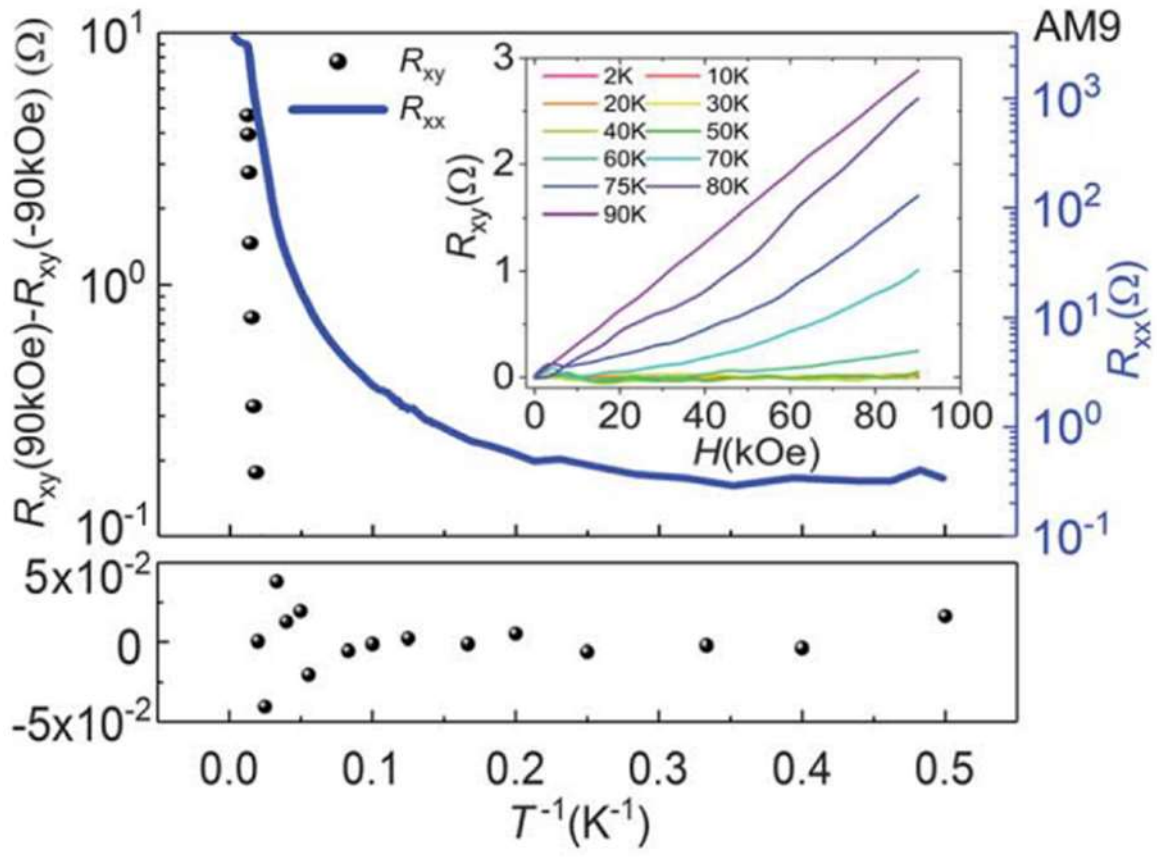

**Figure 20.** Arrhenius plots of Hall resistance (the black dots) and longitudinal resistance (the blue curve) in a representative YBCO film showing the anomalous metal behavior. The Hall

coefficient drops to zero below 50 K within the resolution of measurement instruments. Inset, the Hall resistance with respect to magnetic field at various temperatures. Reprinted from [156].

More interestingly, pronounced magnetoconductance oscillations are observed in nanopatterned YBCO films. Figure 21(a-c) presents the negative change of magnetoconductance $-\Delta G = G_0 - G(H)$ (where $G_0$ is the conductance at zero magnetic field) in three representative films of superconducting state, anomalous metal state and insulating state. Up to four magnetoconductance oscillations are detected with the oscillation period of around 2.25 kOe, which corresponds to superconducting flux quantum $\Phi_0 = \frac{h}{2e}$ (where $e$ is the electron charge and $h$ is the Planck's constant) per unit cell of the nanopattern (the area is around 9200 nm$^2$). The observation of charge-2$e$ quantum oscillations in nanopatterned YBCO films of superconducting state, anomalous metal state and insulating state indicates that the Cooper pairs dominate the transport across the superconductor-anomalous metal-insulator transitions. The oscillations appear just below the onset $T_c$, and the temperature dependence of the oscillation amplitude $G_{osc}$ shows a saturation behavior below 5 K in the films showing anomalous metal state (Fig. 21(d)). The phase coherence length $L_\phi$ could be estimated from $G_{osc}$ by using the formula $G_{osc} = \frac{4e^2}{h}(\frac{L_\phi}{\pi r})^{1.5}\exp(-\frac{\pi r}{L_\phi})$ (where $r$ is half of the center-to-center nanohole spacing) for quasi-particle quantum interference [156, 169]. The temperature dependence of $L_\phi$ (Fig. 21(e)) shows phase coherence length saturation in the low-temperature regime for anomalous metal states. The concurrence of resistance saturation and phase coherence length saturation indicates the prominent role of phase coherence saturation in the anomalous metal state [169-171].

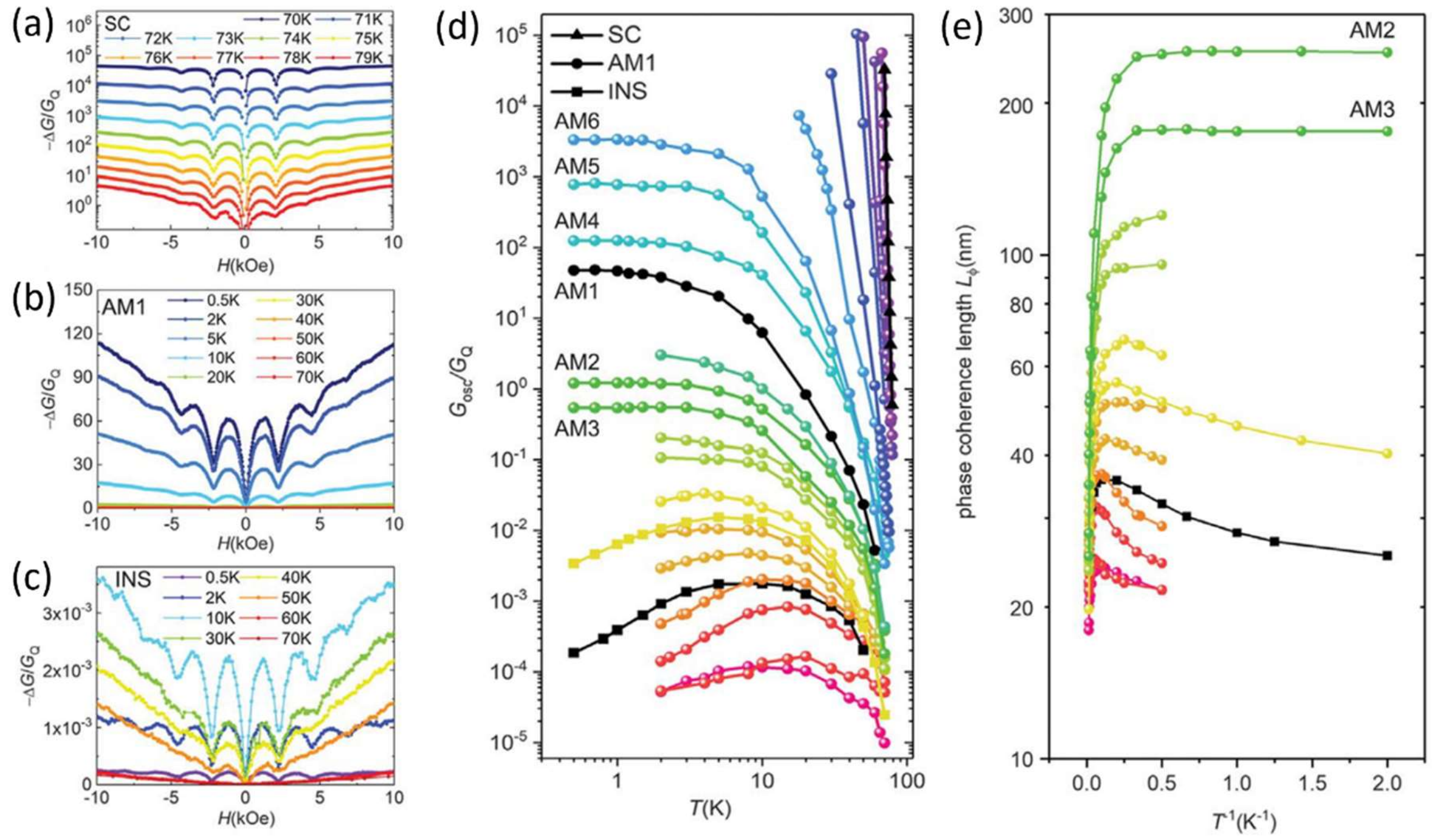

**Figure 21.** (a-c) Magnetoconductance quantum oscillations in nanopatterned YBCO films of (a) superconducting state, (b) anomalous metal state, and (c) insulating state. Up to four magnetoconductance oscillations were detected with a period of around 2.25 kOe, which corresponds to superconducting flux quantum per unit cell of the nanopattern. (d) Temperature dependence of magnetoconductance oscillations amplitude in nanopatterned YBCO films on a logarithmic scale. (e) Phase coherence length derived from magnetoconductance oscillations amplitude plotted as a function of $1/T$. Reprinted from [156].

The vanishing Hall resistance [136] and absence of cyclotron resonance [140, 141] have suggested an emergent particle-hole symmetry in the anomalous metal state. In nanopatterned YBCO films, the observation of charge-$2e$ quantum oscillations demonstrates the bosonic nature of the anomalous metal state, which is considered as solid evidence to end the long-standing debate on whether bosons can exist as a metal [172]. Multiple physical scenarios have been developed to describe the anomalous metal behavior, such as the Bose metal [129, 173, 174], the quantum fluctuation of superconducting order parameter [175-177], the coupling to dissipative heat bath [148, 178, 179], and the composite Fermi liquid [180]. The charge-$2e$ quantum oscillations appear to strongly support the scenarios that charge-$2e$ carriers dominate the transport while rule out the scenarios that only consider charge-$e$ carriers.

## 3.3 Discussion

### 3.3.1 The relation between anomalous metal state and quantum Griffiths singularity

Previous works report the coexistence of QGS and anomalous metal state in several 2D crystalline superconductors (*e.g.*, MBE-grown $PdTe_2$ thin films [109, 161], exfoliated ZrNCl and $MoS_2$ devices [115]), suggesting possible connection between these quantum phenomena. The phase diagram of 4-ML $PdTe_2$ film is summarized in Fig. 22. The boundary of 2D superconductivity (the violet curve) is defined as the critical magnetic field below which zero resistance is detected within the measurement resolution. In the moderate magnetic field regime, the anomalous metal state appears at lower temperatures below $T_{\mathrm{AM}}$ while the thermal creep region appears at higher temperatures above $T_{\mathrm{AM}}$, where $T_{\mathrm{AM}}$ (the light blue curve) is defined as the characteristic temperature when the sheet resistance deviates from the thermally activated behavior and begins to saturate. The thermal creep and the thermal fluctuation are separated by the red curve, which represents the critical field defined by 50% $R_{\mathrm{n}}$. In the high magnetic field and ultralow temperature regime, quantum fluctuation and the quenched disorder rule in the quantum phase transitions, leading to quantum Griffiths state characterized by divergent $z\nu$ along the phase boundary (the brown curve determined from the crossing points of adjacent $R_{\mathrm{s}}(B)$ curves). Remarkably, the system undergoes a superconductor to anomalous metal transition, followed by an anomalous metal to weakly localized metal transition exhibiting QGS in the vicinity of the infinite-randomness quantum critical point $B_c^*$. Similar phase diagram displaying both quantum Griffiths state and anomalous metal state has also been reported in ZrNCl nanodevices [115]. Further experimental and theoretical studies are highly desired to fully understand the relation between the QGS and anomalous metal state.

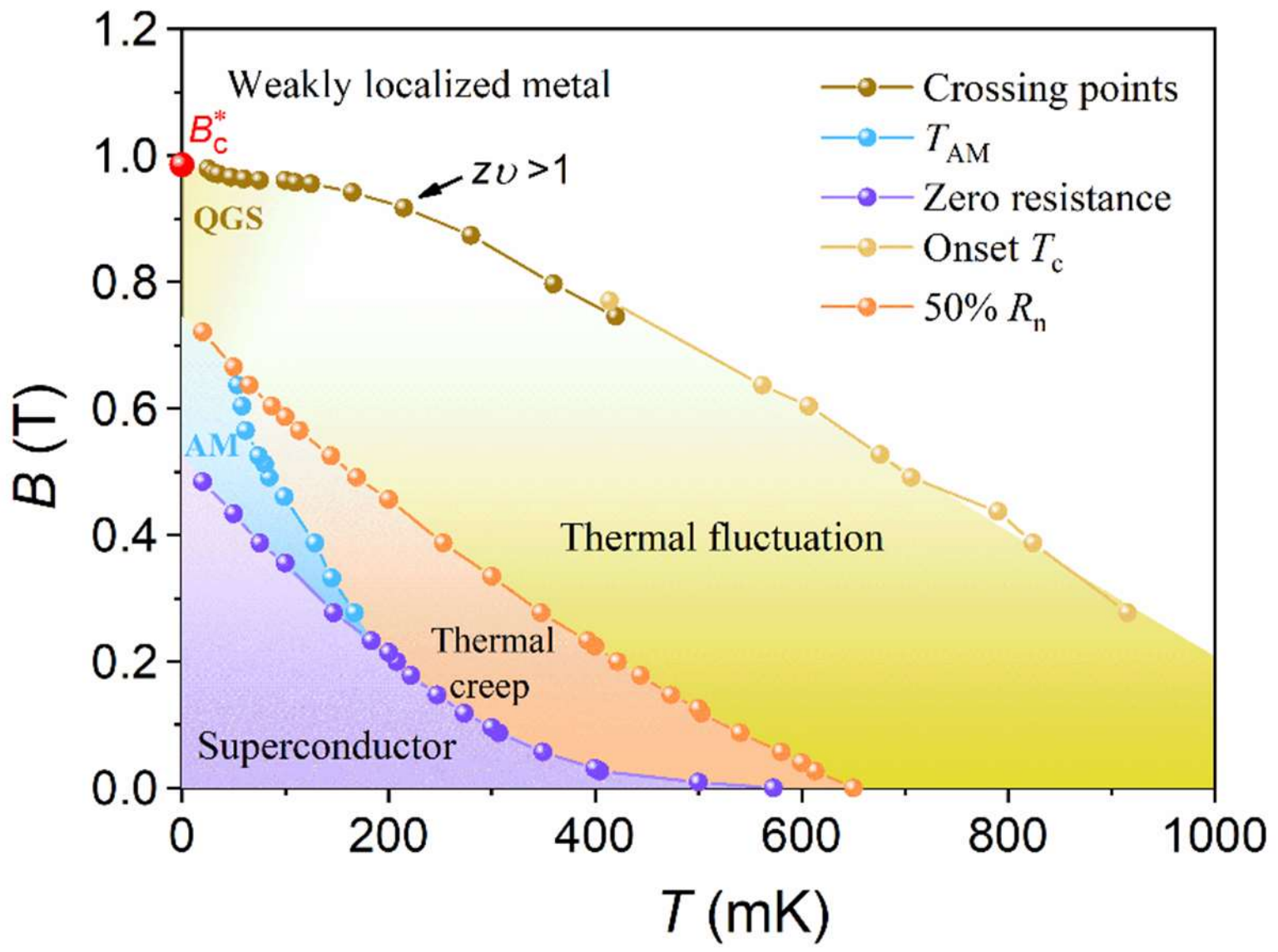

**Figure 22.** Phase diagram of the 4-ML $PdTe_2$ film under out-of-plane magnetic field. At zero temperature, the system undergoes a superconductor to anomalous metal (marked as AM) transition, followed by an anomalous metal to weakly localized metal transition with the infinite-randomness quantum critical point $B_c^*$. QGS characterized by divergent $z\nu$ is revealed at low temperatures along the phase boundary determined from the crossing points of adjacent $R_s(B)$ curves. The data of this phase diagram is from [109, 161].

### 3.3.2 The relation between dissipation and anomalous metal state

In previous studies of conventional superconductors, the anomalous metal state only emerges at ultralow temperatures, which is typically in the hundreds of millikelvin range. High-temperature cuprate superconductors offer the opportunity to study the anomalous metal state with higher onset temperature ($T_{\text{AM}}$ is up to 10 K), but the ratio between the onset anomalous metal temperature and the onset superconducting temperature ($T_{\text{AM}}/T_c^{onset}$) is comparably small [156, 157]. Recently, Li *et al.* reported the observation of the anomalous metal state in ultrathin crystalline FeSe/STO films [32], which stands for the first report of the anomalous metal state in high-temperature superconductors at the 2D limit. The anomalous metal state is detected with record high $T_{\text{AM}}$ up to 20 K and large $T_{\text{AM}}/T_c^{onset}$ ratio (see Fig. 22(a) for $T_{\text{AM}}$ and $T_{\text{AM}}/T_c^{onset}$ in FeSe/STO and other superconducting systems). High $T_{\text{AM}}$ is promising for the exploration of the anomalous metal state using other experimental techniques such as infrared optical conductivity, which are difficult to be performed at ultralow temperatures. The anomalous metal state was also detected in nanopatterned FeSe/STO films, wherein the observation of the charge-2$e$ quantum oscillations demonstrates the bosonic nature of the anomalous metal state [138].

Theoretical understanding of the anomalous metal state is one of the major open issues in condensed matter theory [125]. The origin of resistance in a bosonic system could be attributed to the classical motion or quantum tunneling of vortices [25, 32]. In the investigation of quantum fluctuation of superconducting order parameter, it has been known the coupling to quasiparticle excitations could give rise to the dissipation and act as the source of phase fluctuations [181, 182]. The dissipation significantly changes the dynamics of vortices tunneling [183-186] and further results in observable consequence in quantum phase transitions [178, 187, 188]. The dissipative quantum tunneling of vortices could lead to the phase mode damping [185], which is the key to understand the small saturated resistance of anomalous metal state at low temperatures. In the following, we briefly introduce how the dissipation changes the dynamics of vortices tunneling and gives rise to the experimental features of anomalous metal state [32].

The 2D dissipative quantum XY model provides a microscopic basis to analyze the dynamics of vortices [32, 183]. The effective action at one single Josephson junction is given as (*i.e.*, the resistively shunted Josephson junction model)

$$S = \int d\tau \left\{ \frac{|\dot{\theta}|^2}{2E_c} - E_J Cos[\theta] \right\} + \frac{\gamma}{4\pi} \int d\tau_1 d\tau_2 \left| \frac{\theta(\tau_1) - \theta(\tau_2)}{\tau_1 - \tau_2} \right|^2. \qquad (8)$$

Here, $E_c$ denotes the charge energy, $E_J$ denotes the Josephson energy, and $\gamma = \frac{h}{4e^2 R_L}$ is the dimensionless dissipation strength with link resistance $R_L$. The dynamics of phase $\theta$ can be obtained by mapping to the spin boson model [185], which yields the phase mode damping rate. In the zero-temperature limit, the mobility of vortices derived from the phase mode damping rate saturates to a constant $\mu_v(T \to 0\ \mathrm{K}) \propto \Gamma\left(\frac{\gamma}{2(1-\gamma)}\right) / \Gamma\left(\frac{1}{2(1-\gamma)}\right)$, giving rise to the saturating resistance at low temperatures as an experimental feature of the anomalous metal state [32]. The saturating resistance at low temperatures could be well described in this model, demonstrating that the dynamics of vortices tunneling with ohmic dissipation is critical to the origin of anomalous metal state. Moreover, this model could also explain the experimental observation that the $T_{\mathrm{AM}}/T_c^{onset}$ ratio decreases with increasing normal state resistance $R_{\mathrm{N}}$ in both pristine and nanopatterned FeSe/STO [32]. Noteworthily, the current model is based on the situation in the absence of external magnetic field. Further theoretical studies are still necessary to gain a comprehensive understanding of the anomalous metal state.

### 3.3.3 Bosonic strange metal state

The detection of the bosonic anomalous metal state has challenged the long belief that the bosons can only exist either as superconducting or insulating state in 2D systems when approaching zero Kelvin. Particularly, the recent experiments with high quality filters have confirmed the intrinsic origin of the bosonic anomalous metal state [156]. These findings are beyond the traditional pictures on bosons and hence open up a frontier to explore new quantum states in bosonic systems.

The linear-in-temperature resistance has been reported in correlated fermionic systems, notably heavy fermion materials [60] and the normal state of cuprate and iron-based superconductors [68-71]. This behavior is an experimental feature of the strange metal state, which is believed to originate from the excitations of quantum criticality. Recent experimental results have extended the strange metal phenomenology to bosonic systems. It is revealed in high-temperature superconducting FeSe/STO and YBCO films that the resistance shows a linear temperature dependence in the region between the anomalous metal state and the normal state [32, 189]. The charge-$2e$ quantum oscillations are detected in the linear-in-temperature resistance region below onset $T_c$ of the nanopatterned samples, and simultaneously the Hall coefficient is observed to drop and vanish with decreasing temperature, indicating the transport is dominated by the Cooper pairs instead of unpaired electrons [32, 189]. These results mark a bosonic strange metal state.

The bosonic strange metal state in the moderate temperature regime may be attributed to the thermal-smeared quantum tunneling of vortices with ohmic dissipation. In the dissipative quantum XY model, the temperature dependence of resistance could be derived considering the dissipation strength and the phase mode damping rate [32]

$$R_S = \frac{h}{4e^2}\frac{E_C}{4k_B T_C^{onset}}\frac{\sqrt{\pi}}{2}\frac{\Gamma(\overline{\gamma})}{\Gamma(\overline{\gamma}+1/2)}\left(\frac{\pi k_B T}{E_J}\right)^{2\overline{\gamma}-1}. \qquad (9)$$

Here, $\overline{\gamma} = \frac{h}{4e^2\overline{R_L}}$ is the mean dissipation strength with the mean link resistance $\overline{R_L}$, $E_c$ and $E_J$ denote the charge energy and the Josephson energy, respectively. It is expected that the linear-in-temperature resistance becomes remarkable when the mean link resistance $\overline{R_L}$ is close to the quantum resistance $R_Q = \frac{h}{4e^2}$ (*i.e.*, $\overline{\gamma} = 1$). Noteworthily, the bosonic strange metal state is observed in a wide range of parameters in both FeSe/STO and cuprate superconducting systems [32, 189], calling for further theoretical considerations.

## 4. Conclusion and outlook

### 4.1 Conclusion

We have reviewed the recent progress on quantum phase transitions in 2D superconducting systems, focusing on the QGS and anomalous metal state. In the investigation of quantum phase transition between a superconducting and a weakly localized metallic state in ultrathin Ga films, a continuous line of crossing points is identified in the magnetoresistance isotherms and a divergent critical exponent is revealed approaching the quantum critical point [14], which are distinct from the single crossing point and constant critical exponent as shown in early studies of SIT/SMT. This new quantum phase transition behavior observed in 2D superconducting films is called as QGS. Since then, the QGS has been widely reported in diverse 2D superconducting systems, which not only demonstrates the universality of QGS but also reveals the profound influence of quenched disorder on quantum phase transitions.

Although possible signature of intervening metallic state had been reported in early studies of 2D superconducting systems, the existence of intrinsic anomalous metal state was unclear due to the low critical temperature and the influence of external high frequency noise and measurement current. In the SIT of nanopatterned high-temperature superconducting YBCO films, resistance saturation in the low temperature regime is revealed as experimental evidence of the anomalous metal state [156]. The resistance saturation remains almost unchanged with or without high-quality filters, indicating the intrinsic anomalous metal state. Moreover, the observed quantum oscillations of Cooper pairs indicate the bosonic nature of the anomalous metal state, which is considered as solid evidence to end the long-standing debate on whether bosons can exist as a metal [172]. With the high-quality filters and the small measurement current within the linear I-V regime, the resistance saturation down to ultralow temperature regime has also been detected in MBE-grown crystalline $PdTe_2$ thin films [161] and exfoliated crystalline $TaSe_2$ nanodevices [152] as well as granular films of $In/InO_x$ composite [146, 147]. These findings provide reliable evidences on the existence of intrinsic anomalous metal state in 2D superconducting systems.

### 4.2 Outlook

**4.2.1 Measurements at lower temperatures.** As illustrated in earlier sections, low temperature is an essential requirement for the study of quantum ground states and quantum phase transitions. Dilution refrigerators are commonly utilized to reach ultralow temperature down to 10 mK range, which is crucial to reveal the resistance saturation of the anomalous metal state and the divergent critical exponent of QGS. Nowadays adiabatic demagnetization technique has been applied to create extremely low temperature environment for condensed matter experiments. It is expected that the extremely low temperature down to 1 millikelvin or lower would not only provide more solid evidence for the QGS and the anomalous metal state but also offer promising opportunity to uncover new quantum states and quantum phase transitions approaching zero Kelvin. For instance, benefiting from the ultralow-temperature magnetic and calorimetric measurements using adiabatic demagnetization refrigerator, a heavy-electron superconducting phase has been revealed at 2 mK in $YbRh_2Si_2$ in the vicinity of antiferromagnetic quantum critical point [190].

**4.2.2 Beyond 2D superconductors.** So far, the investigations of QGS and anomalous metal state in superconductors mainly focus on 2D systems. 1D superconductors provide valuable platform to study quantum phase transitions since 1D systems tend to have stronger effect of disorder and quantum fluctuations compared to 2D systems. Quantum phase transitions might also occur in 3D superconductors with strong fluctuation effect. Recently, the anomalous metal state has been revealed in highly compressed bulk titanium [191]. The detections of anomalous metal states in 1D and 3D superconductors are expected to deepen the understanding on dissipative quantum tunneling of vortices in systems with different dimensionalities. Furthermore, the observations of QGS in 1D and 3D superconductors are highly desired to demonstrate the relation between the critical behavior and dimensionality.

**4.2.3 Other experimental techniques.** In this review, we have demonstrated the experimental evidences of quantum phase transitions in 2D superconducting systems obtained by electrical transport measurements. Recently, other experimental techniques such as thermal transport measurements [142, 192, 193] and nonlinear optical measurements [194] have been utilized in the study of quantum phase transitions in superconductors. For instance, Ienega *et al.* uncovered superconducting fluctuations in the vicinity of quantum critical point in amorphous $Mo_xGe_{1-x}$ thin films via the measurements of Nernst effect [142]. Besides, Alcalà *et al.* reported the nonlinear optical signatures of quantum phase transitions in the high-temperature superconducting YBCO films [194]. These experimental techniques provide powerful methodology and new insights in the quantum phase transitions. It is highly anticipated the applications of new experimental techniques on 2D crystalline superconductors, particularly the techniques to image Cooper pair formation and vortices motion (such as scanning tunneling microscopy [195, 196] and scanning superconducting quantum interference device [197]), would offer a promising route to understanding the microscopic picture of quantum state and quantum phase transitions including the anomalous metal state, the bosonic strange metal state and the QGS.

**4.2.4 Exploring new quantum phase transitions.** Research in quantum phase transitions mainly focus on continuous phase transitions from an ordered ground state (with nonzero order parameter) to a quantum disordered state (with zero order parameter), covering SIT/SMT, ferromagnetic to paramagnetic transitions, and ferroelectric to paramagnetic transitions. In such continuous quantum phase transitions, the quantum critical behavior could be described in the LGW paradigm [58]. It has been theoretically proposed that continuous quantum phase transition could also occur between two ordered ground states with different broken symmetries [57, 198, 199], which is beyond the LGW paradigm. The study of deconfined quantum critical point may shed light on experimental puzzles in strongly correlated systems including cuprate superconductors and also offer opportunity to explore fractional excitations and emergent symmetries [57, 200]. However, the experimental realization of the proposed deconfined quantum critical point is very challenging. Recently, high-pressure NMR measurements provide possible experimental indication of deconfined quantum critical point that connects two different ordered ground states in the quantum magnet $SrCu_2(BO_3)_2$ [201]. The 2D superconductors with strong electron correlations may also offer a platform to explore new quantum phase transitions beyond the LGW paradigm.

**4.2.5. Disorder and fluctuations in complex physical systems.** Disorder and fluctuations play an essential role in the quantum phase transitions. A wide variety of theoretical models have been developed to understand the role of disorder and fluctuations in condensed matter systems. For instance, spin glass model was developed in 1970s to describe the phase transition in disordered magnetic systems. In 1979, Parisi demonstrated that the replica symmetry breaking could be used to solve a spin glass model [202]. The hidden pattern discovered in replicas has been extended to understand and describe the disordered complex systems in many different areas such as turbulent systems, neural networks and machine learning. Giorgio Parisi was awarded half of the 2021 Nobel Prize in Physics “for the discovery of the interplay of disorder and fluctuations in physical systems from atomic to planetary scales". It is anticipated the investigations of disorder and fluctuations in quantum phase transitions (*e.g.*, the rare regions in QGS) of 2D superconductors may also inspire the studies in various research areas.

**Data availability statement**

No new data were created or analyzed in this study.

**Acknowledgments**

We acknowledge Haiwen Liu for fruitful discussion on this review. We thank Hai-Long Fu, Jun Ge, Yanrong Li, Xi Lin, Chong Liu, Haiwen Liu, Shuaihua Ji, Xu-Cun Ma, Jiacai Nie, Shichao Qi, Shengchun Shen, Jian Sun, Yi Sun, James M. Valles Jr., Fa Wang, Lili Wang, Pengjie Wang, X. C. Xie, Ying Xing, Jie Xiong, Jimmy Xu, Qi-Kun Xue, Chao Yang, Pu Yang, Hong Yao, Hui-Min Zhang, Kun Zhao and other collaborators for collaborations related to the subject of this review. This work was supported by the National Key Research and Development Program of China (Grant No. 2018YFA0305604, No. 2022YFA1403103), the National Natural Science Foundation of China (Grant No. 11888101), the Innovation Program for Quantum Science and Technology (2021ZD0302403), the National Natural Science Foundation of China (Grant No. 12174442), Young Elite Scientists Sponsorship Program by BAST (No. BYESS2023452), the Fundamental Research Funds for the Central Universities and the Research Funds of Renmin University of China (Grant No. 22XNKJ20).